\documentclass[sigconf, nonacm]{acmart}
\usepackage{graphicx}
\usepackage{balance}  
\usepackage{booktabs} 
\usepackage{multirow}
\usepackage{url}
\usepackage{mindatadog}

\newcommand\availabilityurl{https://github.com/datadog/ssss-paper-code}
\begin{document}
\title{Sampling Space-Saving Set Sketches}

\author{Homin K. Lee}
\affiliation{%
  \institution{Datadog}
  \city{New York City}
  \country{USA}
}
\email{homin@datadog.com}

\author{Charles Masson}
\affiliation{%
  \institution{Datadog}
  \city{New York City}
  \country{USA}
}
\email{charles@datadog.com}

\begin{abstract}
Large, distributed data streams are now ubiquitous. High-accuracy
\emph{sketches} with low memory overhead have become the de
facto method for analyzing this data. For instance, if we wish to group data by
some label and report the largest counts using fixed memory, we need to turn to
mergeable heavy hitter sketches that can provide highly accurate
approximate counts. Similarly, if we wish to keep track of the number of
distinct items in a \emph{single} set spread across several streams using fixed
memory, we can turn to mergeable count distinct sketches that can
provide highly accurate set cardinalities.

If we were to try to keep track of the cardinality of \emph{multiple} sets and
report only on the largest ones, maintaining individual count distinct
sketches for each set can grow unwieldy, especially if the number of sets is not
known in advance. We consider the natural combination of the heavy hitters
problem with the count distinct problem, the \emph{heavy distinct hitters}
problem: given a stream of $(\ell, x)$ pairs, find all the labels $\ell$ that
are paired with a large number of distinct items $x$ using only constant memory.

No previous work on heavy distinct hitters has managed to be of practical use in
the large, distributed data stream setting. We propose a new algorithm, the
Sampling Space-Saving Set Sketch, which combines sketching and sampling
techniques and has all the desired properties for size, speed, accuracy,
mergeability, and invertibility. We compare our algorithm to several existing
solutions to the heavy distinct hitters problem, and provide experimental
results across several data sets showing the superiority of the new sketch.

\end{abstract}

\maketitle

\begingroup
\renewcommand\thefootnote{}\footnote{\noindent
This work is licensed under the Creative Commons BY-NC-ND 4.0 International License. Visit \url{https://creativecommons.org/licenses/by-nc-nd/4.0/} to view a copy of this license. For any use beyond those covered by this license, obtain permission by emailing the authors. Copyright is held by the owner/author(s). \\
}\addtocounter{footnote}{-1}\endgroup

\ifdefempty{\availabilityurl}{}{
\vspace{.3cm}
\begingroup\small\noindent\raggedright\textbf{Artifact Availability:}\\
The source code, data, and/or other artifacts have been made available at \url{\availabilityurl}.
\endgroup
}

\section{Introduction}

When analyzing distributed data, it is crucial to be able to report accurate
summary statistics~\cite{BJPM-2016}. Some statistics, such as the average, are
trivial to compute in a distributed system. Others, such as
quantiles~\cite{MRL-2019}, require more sophisticated sketches of the data.

We consider a particularly difficult case, the \emph{heavy distinct hitters}
problem: given a stream of $(\ell, x)$ pairs, find all the labels $\ell$ that
are paired with a large number of distinct items $x$ (which we call \emph{heavy
sets}) using only constant memory.
The set of labels are not known in
advance.
The definition of ``large'' can either be formulated as any set whose
cardinality is over some absolute threshold, or we can define it to be the top
$k$ sets by cardinality, e.g., queries of the form:

\begin{equation}
  \begin{array}{l}
    \verb!SELECT L, COUNT(DISTINCT X) AS C! \\
    \verb!GROUP BY L!\\
    \verb!ORDER BY C DESC!\\
    \verb!LIMIT K!\\
  \end{array} \label{query}
\end{equation}

This problem has also been referred to as the \emph{distinct heavy hitter}
problem~\cite{FABCS-2017} as well as the \emph{superspreader}
problem~\cite{VSGB-2005} in the context of network security. The network
security use cases include detecting source IP addresses that make an unusually
high number of distinct connections within a short time, and detecting random
subdomain distributed denial of service (DDoS) attacks on the Domain Name System
(DNS) infrastructure~\cite{Weber-2014} (where the label is the domain name and
the items are the sub-domains). Other use cases include web applications that
want to track the number of distinct users that have clicked on a
recommendation, or advertisers that want to track the ad campaigns seen by the
largest number of distinct users~\cite{MAA-2008}.

The use case that motivates this work comes from agent-based infrastructure
monitoring.  As computing has moved to a distributed, containerized,
micro-service model, many organizations use container orchestration to run
thousands of pods (tightly coupled collections of
containers)~\cite{Datadog-orchestration}.  Effectively being able to administer
and operationalize such a large fleet of machines requires the ability to
monitor, in near real-time, data streams coming from multiple transitory
sources~\cite{BJPM-2016}. A common solution is to use agents to collect the
necessary observability metrics, and it is in this setting where no previous
work on heavy distinct hitters has managed to be of practical use.

\begin{figure}
  \centering \includegraphics[width=.5\textwidth]{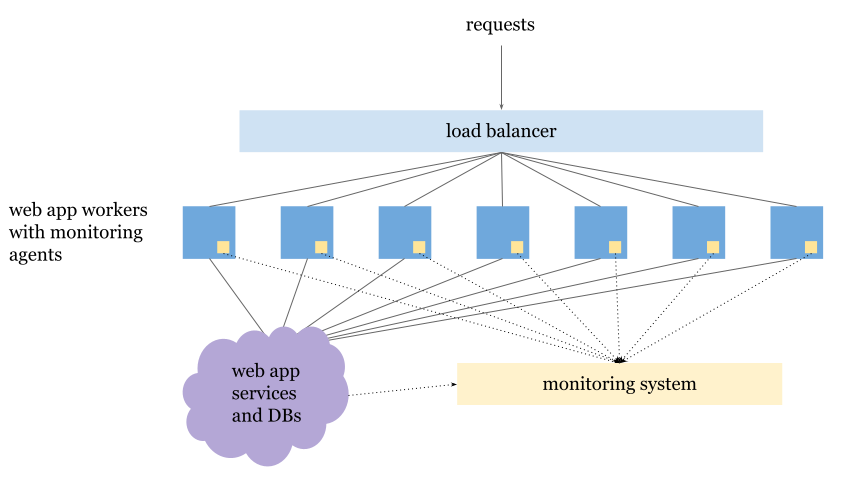}
  \caption{A distributed web application, with each pod sending metrics to
    the monitoring system.}
  \label{fig:monitoring}
  \Description[monitoring]{sketch of infrastructure diagram for a distributed web
    application, with each pod sending metrics to the monitoring system}
\end{figure}

Figure~\ref{fig:monitoring} shows a typical example of a web application backed
by a distributed system. Every machine in the distributed system runs a small
agent that reports statistics to the monitoring system.  When agents run on
machines it is crucial that the agents consume a predictably small amount of
compute, memory, and network usage so that the machines can reliably run their
primary workloads as production services. The necessary properties for a sketch
for the \emph{heavy distinct hitter} problem to be practical in this setting
are:

\begin{itemize}
\item {\bf Single Pass.} The sketch gets one look at each element over any data
  stream.
\item {\bf Accuracy.} We want our sketch approximations to return values that
  are close to the actual set cardinalities. We will consider the normalized
  absolute error, defined in Section~\ref{subsec:metrics}.
\item {\bf Constant Memory Size.} The memory footprint of the sketch should be
  independent of the amount of data processed by the sketch.
    For a sketch to be
  viable in the agent monitoring model, the memory size should be on the order
  of megabytes.
\item {\bf Insertion Speed.} We want our sketches to be able to process
  high-throughput data streams without falling behind.
\item {\bf Mergeability.} When aggregating over distributed data streams, we
  want to be able to sketch each data stream separately and then merge the
  sketches to provide answers over the entire data set~\cite{ACHPWY-2012}
  without losing too much accuracy.
\item {\bf Invertibility.} The sketch itself needs to be able to return
  the labels of all the heavy distinct hitters~\cite{THL-2020}. Non-invertible
  sketches are able to return the approximate cardinality for any label provided
  at query time, but do not store any labels themselves.
\item {\bf Query Speed.} The sketch should have fast query response times. An
  increasingly common pattern is to sketch data streams every few seconds, and
  report statistics as time series \cite{MassonWatt-2019}.
\end{itemize}

These properties are also necessary when a distributed
database~\cite{GormleyTong-2015, Datadog-husky} responds to queries such as
(\ref{query}). Gathering all the relevant data in a centralized node to give an
exact answer is often infeasible for large data. An accurate constant-size
sketch allows for each leaf node to return partial aggregates, which are then
merged by the query engine. The constant-size is important as databases need to
be able to respond to many queries at once using a finite amount of
resources. Note that non-invertible sketches by definition can not provide
responses to queries such as (\ref{query}).

We present the Sampling Space-Saving Set Sketch, the first sketch with all of
the above properties. In Section~\ref{sec:model} we formally define the heavy
hitter, count distinct, and heavy distinct hitter problems. In
Section~\ref{sec:algo1} we define an initial version of the algorithm and prove
bounds on its accuracy. We present the final version of the algorithm in
Section~\ref{sec:algo2}. Finally, in Section~\ref{sec:exp} we present our
experimental results showing that Sampling Space-Saving Set Sketches provide
superior accuracy, memory usage, and throughput than existing sketches.

The source code of our prototype along with the code for running our experiments
is freely available online.

\section{Related Work}\label{sec:model}
Given a data stream $x_1\through x_m$ of elements coming from a universe of size
$n$, let $f_i = |\{ k : x_k = i \}|$ be the frequency of element $i$ so that
$\sum_i f_i = m$.

\subsection{Count Distinct}
The \emph{count distinct} problem, also known as the \emph{set cardinality}, or
\emph{distinct elements}, or $F_0$ problem is to output $d = |\{i : f_i > 0\}|$,
the number of distinct elements seen in the stream.

There are a variety of sketch algorithms that return approximate answers while
minimizing the amount data stored, including Probabilistic
Counting~\cite{FlajoletMartin-1983}, HyperLogLog~\cite{FFGM-2007}, Adaptive
Sampling~\cite{Flajolet-1990}, $K$'th Minimum Value~\cite{BJKST-2002},
Alpha~\cite{DLRT-2016}, and Compressed Probabilistic
Counting~\cite{Lang-2017}. Each sketch algorithm has an \textsc{Insert} method
for adding elements, and a \textsc{Distinct} method for estimating the
cardinality of the set of elements added so far.

Count distinct sketch algorithms are required to output an estimate of the
cardinality of the set with relative error $\eps$, with probability at least
$1-\delta_c$. \Ie if the exact number of distinct elements is $d$, return
$\tilde{d}$ such that $(1-\eps) d \leq \tilde{d} \leq (1+ \eps)d$.

For many practical applications, the standard one-shot guarantee is too weak, as
the sketch can be queried several times while it processes a stream. A
count distinct algorithm is said to have a \emph{strong-tracking} guarantee
\cite{HTY-2012, Blasiok-2018} if it reports an accurate estimate after each
\textsc{Insert}:
\[\Pr[\forall k, (1-\eps) d^{(k)} \leq \tilde{d}^{(k)} \leq (1+ \eps)d^{(k)} ] \geq 1-\delta_c,\]
where $d^{(k)}$ is the number of distinct elements seen at time step $k$.

The size of the sketches are functions of $\eps$, $\delta_c$, and $n$, where the
optimal bound is $\Theta(\eps^{-2}\log(1/\delta_c) + \log n)$ for the one-shot
version \cite{KNW-2010}, and $\Theta(\eps^{-2}(\log\log n + \log(1/\delta_c)) +
\log n)$ for the strong-tracking version \cite{Blasiok-2018}.  In addition to
the size, the time complexity of \textsc{Insert} is very important as it gets
called on each item in the stream.

Finally, we will only be considering \emph{monotonic} count distinct
sketches. That is we only consider sketches whose estimate of the set
cardinality can only increase as more elements are added to the sketch. While
the error guarantee allows for a count distinct sketch to be non-monotonic, most
count distinct sketches in the literature are monotonic, as non-monotonic
behavior would be very counter-intuitive.

\subsection{Heavy Hitters}
The heavy hitter problem, with parameter $s$, is to output all elements with
frequency $f_i > m/s$, where $m$ is the number of elements in the stream. An
$s$-counter heavy hitter algorithm keeps at most $s$ counters and is guaranteed
to return for every item an estimate $\tilde{f}_i$ such that $\abs{\tilde{f}_i -
  f_i} \leq m / s$

The are a variety of sketch algorithms for this problem, including
Misra-Gries~\cite{MisraGries-1982}, Space-Saving~\cite{MAA-2005},
Count~\cite{CCF-2002}, and Count-Min~\cite{CormodeMuthukrishnan-2004}. The size
of the sketches are functions of $s$, $m$, and $n$ (as well as a failure
probability parameter $\delta$ if the sketch is randomized).

For insertion-only streams, the Space-Saving sketch has been found to have clear
benefits over the other sketches~\cite{CormodeMuthukrishnan-2010}. A
Space-Saving sketch of size $s$ stores $s$ $(\ell, count)$ pairs. It is
initialized with the first $s$ distinct labels seen along with their exact
counts. If a subsequent item is already in the sketch, the count is
updated. Otherwise, the pair with the smallest count will have its label
replaced with the new label, and then the count will be incremented. The
insertion time is dominated by the time to find the minimum count label, where
various strategies are possible (including that of maintaining a min-heap).

\subsection{Heavy Distinct Hitters}\label{subsec:hdh_previous_work}
The \emph{heavy distinct hitters} problem combines the \emph{count distinct}
problem with the \emph{heavy hitters} problem.  Given a data stream of elements
each with a label, $(\ell_1, x_1)\through (\ell_m, x_m)$, let
$f_{\ell,x} = |\{ k : \ell_k = \ell, x_k = x \}|$ be the frequency of item $x$
with label $\ell$, and let $d_\ell = |\{x : f_{\ell,x} > 0\}|$ be the number of
distinct items with label $\ell$. The heavy distinct hitters problem, with
parameter $s$, is to output all labels with $d_\ell > m/s$, and for every label,
return an estimate $\tilde{d}_\ell$ such that $\abs{\tilde{d}_\ell - d_\ell} \leq m(1 / s + \eps)$
with probability at least $1-\delta$.
The labels can be arbitrary, and the set of
labels are not known in advance.

Initial papers tackled the heavy distinct hitter problem by just keeping a set
for every label~\cite{Roesch-1999, Plonka-2000}, which was then improved to
keeping a count distinct sketch for every label~\cite{EVF-2003}. The first paper
to consider the heavy distinct hitter problem holistically (and which introduced
the terminology) was by Venkataraman et al.~\cite{VSGB-2005} whose
sampling-based solution was improved in subsequent work~\cite{KMK-2007,
  CJCBZ-2009, Locher-2011}. The sampling based approaches require a predefined
sampling rate, which is impractical in the distributed stream setting, and it
requires knowing the total length of the stream in advance to match the required
bounds.

A related problem is the \emph{many distinct count} problem, which requires that
the sketch accurately answer the count distinct problem for multiple labels,
whether or not their corresponding sets are heavy. A key difference with the
heavy distinct hitter problem is that the sketch itself is not required to hold
any labels, and thus is not \emph{invertible}. Note that a sketch for the heavy
distinct hitter problem can be trivially modified to be a many distinct count
sketch by outputting either the minimum cardinality or zero for any label not in
the sketch.

The state-of-the-art sketch for the many distinct count problem is
Count-HLL~\cite{Ting-2019}, which uses a Count-Min sketch whose columns are
replaced with the registers of HLL sketches. Previous work includes the CMFM
sketch~\cite{CHLBK-2009}, a Count-Min sketch whose counters are replaced with
Probabilistic Counting sketches~\cite{FlajoletMartin-1983}, and
vHLL~\cite{XCCL-2015}, which is similar to Count-HLL, but has a different
hashing scheme and final cardinality estimator.

Many other non-invertible sketches exist~\cite{ZKX-2005, YLCP-2011, LCLZQ-2013},
and the general method for turning them into invertible sketches is to
separately keep a set of all the labels in the stream, and to then iterate
through the \emph{entire} set of labels when finding the heavy hitters.
Some approaches work to reduce the search space~\cite{WGQH-2011, LQGL-2016}.

Invertible sketches for the heavy distinct hitter problem include Distinct-Count
Sketch~\cite{GGRS-2007}, Count-Min-Heap~\cite{CormodeMuthukrishnan-2005},
OpenSketch~\cite{YJM-2013}, FastSketch combined with count distinct
sketches~\cite{LCG-2016}, and SpreadSketch~\cite{THL-2020}, which was shown by
the authors to be superior to the previous four sketches in memory, insertion time
and query time. Much like several of the solutions for the count distinct
problem, SpreadSketch uses a Count-Min sketch, and replaces the counters in the
array with a count distinct sketch. To make the sketch invertible, they add a
candidate heavy distinct hitter label along with a rough estimate of that
label's cardinality in each array entry.

Finally, we note the invertible algorithm idwsHH~\cite{FABCS-2017} by Feibish et
al. They start by generalizing the sampling algorithms for the heavy distinct
hitters problem, and importantly, they add two twists that ends up giving them
something very similar to the Space-Saving sketch for heavy hitters. First, they
fix the number of count distinct sketches they keep. Second, when there is a
label that is not in the sketch, they drop the count distinct sketch with the
highest random hash of label and item only if the new hash value $h(\ell_i,
x_i)$ passes a sampling threshold. This threshold gets dynamically updated every
time it is passed by a new $(\ell_i, x_i)$ pair, and thus the threshold does not
need to be set manually. The downsides of their sketch are that it is not
mergeable, and that it has poor accuracy. The poor accuracy partially stems from
their use of a count distinct sketch that is not particularly robust. More
significantly, the poor accuracy comes from many heavy labels being missing
entirely, as each label has an independent failure probability.

\section{Space-Saving Set Sketches}\label{sec:algo1}

Algorithm~\ref{alg:osss} is a natural extension of the Space-Saving algorithm
for the heavy hitters problem to the heavy distinct hitters problem (we call it
the Space-Saving Sets algorithm). It keeps up to $s$ count distinct sketches
along with corresponding offset values.  We call each (count distinct sketch,
offset value) pair, a \emph{counter} to emphasize the similarity to the
Space-Saving algorithm.

For the first $s$ distinct labels, the sketch just adds items to their
corresponding count distinct sketches. When a new label arrives, the sketch
creates a new counter for the label, finds the minimum existing counter and
records its value into the offset for the new counter, and then removes the
minimum existing counter.

\begin{algorithm}
  \Init{(s)}{ $S \leftarrow $ empty associative array of strong-tracking,
    monotonic count distinct sketches with \textsc{NewSketch}, \textsc{Insert},
    \textsc{Distinct} methods, accuracy parameter $\eps$ and failure parameter
    $\delta_c$\;

    $O \leftarrow $ empty associative array of
    offset values\;
  }

  \Fn{\textsc{Query}($\ell$)}{
    \If{$\ell \in S$} {
      \Return $S[\ell].\textsc{Distinct()} + O[\ell]$\;
    } \Else {
      \Return $\min_{j\in S} S[j].\textsc{Distinct}()$\;
    }
  }

  \Fn{\textsc{Remove}($\ell$)}{
    $S.\textsc{Remove}(\ell)$\;
    $O.\textsc{Remove}(\ell)$\;
  }

  \Fn{\textsc{Insert}($\ell$, x)}{
    \If{$\ell \not\in S$;}{
      \If{$|S| < s$;}{
        $S.\textsc{Add}(\ell,\textsc{NewSketch()})$\;
      } \Else {
        \tcc{save the smallest counter value to the new label offset}
        $y \leftarrow \arg \min_{j\in S}$ $\textsc{Query}(j)$\;
        $O[\ell] \leftarrow \textsc{Query}(y)$\;
        $\textsc{Remove}(y)$\;
        $S.\textsc{Add}(\ell,\textsc{NewSketch()})$\;
      }
    }
    $S[\ell]$.\textsc{Insert}$(x)$\;
  }

  \caption{\textsc{Space-Saving-Sets($s$)}\label{alg:osss}}
\end{algorithm}

We will first show a couple properties of the minimum value count distinct sketch
in the algorithm.

\begin{lemma}\label{lem:min-props}
  After $m$ insertions, let $d_\ell$ be
  the number of distinct elements $x_i$ with label $\ell$.  Let $\alpha :=
  \min_{j\in S} S[j].\textsc{Distinct}()$, and let $\abs{\{\ell : d_\ell > 0\}}
  > s$. Then, for Algorithm~\ref{alg:osss}:

  \begin{enumerate}
  \item $\alpha$ never decreases as entries are inserted;
  \item $\alpha \leq (m/s)(1+\eps)$;
  \item $\textsc{Query}(\ell) \leq d_\ell(1+\eps) + \alpha$; and
  \item if $\ell\not\in S$, then $d_\ell \leq \alpha/(1-\eps)$.
  \end{enumerate}
  where the last three properties hold with probability at least $1-\delta_c$ by
  the strong-tracking property of the count distinct sketch, and $\eps$ is its
  accuracy guarantee.
\end{lemma}
\begin{proof}
  The first property comes from the use of \emph{monotonic}
  count distinct sketches, and the recording of the value of a counter into an
  offset before deleting it.

  For the second property, note that every one of the $m$ items is inserted into
  a count distinct sketch exactly once. Thus, the total sum of all the counters
  is at most $m(1+\eps)$ by the strong-tracking sketch guarantee, and the
  smallest value can be at most $(m/s)(1+\eps)$.

  For the third property, we upper-bound each component of the counter
  separately. The count distinct sketch itself has at most $d_\ell$ items
  inserted into it, and it has at most $\eps$ relative-error by the sketch
  guarantee.  The offset for any counter is always assigned the value of the
  minimum counter when that counter is created. By the first property, the
  minimum value only ever increases until it reaches $\alpha$.

  For the final property, consider the final time the counter for $\ell$ gets
  removed from $S$. At the time of removal, its value (let us call it $\alpha'$)
  is the minimum among all the counters. Then we can bound $\alpha'$ by:
  \[ \alpha \geq \alpha' \geq  d_\ell (1-\eps). \]
  Where the upper-bound comes from the first property, and the lower-bound comes
  from the sketch guarantee.
\end{proof}

Now we are ready to prove the main theorem.

\begin{theorem}\label{thm:main}
  For any label $\ell$, let $d_\ell$ be the number of distinct elements $x_i$
  with label $\ell$ in the stream.  After $m$ insertions to
  Algorithm~\ref{alg:osss} with parameter $s > 4$, and $\eps < 1/2$,
  $\abs{\textsc{Query}(\ell) - d_\ell} \leq m(1/s + (1+1/s)\eps)$, with
  probability at least $1-\delta_c$.
\end{theorem}
\begin{proof}
  If $\ell\not \in S$, by properties 2 and 4 from Lemma~\ref{lem:min-props}, we
  have that the error is at most:
  \begin{eqnarray*}
  |\alpha - d_\ell| &
  \leq &
  \max(\alpha, d_\ell)
  \leq
  \frac{\alpha}{1-\eps}
  \leq
  \Paren{\frac{m}{s}}\Paren{\frac{1+\eps}{1-\eps}}\\
  & = &
  \Paren{\frac{m}{s}}\Paren{1+ \frac{2\eps}{1-\eps}}
  \leq
  \Paren{\frac{m}{s}} + m\eps,
  \end{eqnarray*}
  where the last inequality holds as $s > 2/(1-\eps)$.

  If $\ell\in S$, by properties 2 and 3 from Lemma~\ref{lem:min-props}, we have
  that the error is at most
  \[
  d_\ell \eps + \alpha
  \leq
  d_\ell \eps + \Paren{\frac{m}{s}}\Paren{1+ \eps}
  \leq
  \Paren{\frac{m}{s}} + \Paren{1 + \frac{1}{s}}m\eps,
  \]
  where the last inequality comes from bounding $d_\ell \leq m$.
\end{proof}

Let us consider some examples for which Theorem~\ref{thm:main} is tight.  If the
stream of elements $(\ell, x_i)$ only ever includes one label, then only one
counter is created. If each $x_i$ is unique, then all the error comes from the
count distinct sketch, and so we have an upper-bound of $m\eps$ on the error.

If the stream of elements $(\ell_i, x_i)$ only has a single element $x$, but the
$\ell_i$'s cycle through the same $s+1$ labels repeatedly, then every insert after
the first $s$ will delete an existing counter.  If the deletions happen
sequentially, then after $m$ inserts, every counter will have offsets of $m/s$.
So instead of reporting the correct count $d_j=1$, the sketch can return values
as large as $(m/s)(1+\eps)$.

Much like the Space-Saving algorithm, the insertion time is dominated by the
time to find the minimum cardinality label, where various strategies are
possible (including that of maintaining a min-heap).

\section{Sampling Space-Saving Set Sketches}\label{sec:algo2}

We will make a series of modifications to Algorithm~\ref{alg:osss} so that the
worst-case bound of Theorem~\ref{thm:main} continues to hold, while improving
the practical performance of the sketch.

\subsection{Recycling the Count Distinct Sketches}

In the \emph{recycling} variant of SSS (Algorithm~\ref{alg:rsss}), instead
of recording the value of the minimum counter into an offset, we simply reuse
the counter. (We only specify the modified \textsc{Insert} method. The other
methods are modified to remove the unused offset values.)

\begin{algorithm}
  \Fn{\textsc{Insert}($\ell$, x)}{
    \If{$\ell \not\in S$;}{
      \If{$|S| < s$;}{
        $S.\textsc{Add}(\ell,\textsc{NewSketch()})$\;
      } \Else {
        \tcc{use the smallest counter as the counter for the new label}
        $y \leftarrow \arg \min_{j\in S} S[j].\textsc{Distinct}()$\;
        $S.\textsc{Add}(\ell, S[y])$\;
        $S.\textsc{Remove}(y)$\;
      }
    }
    $S[\ell]$.\textsc{Insert}$(x)$\;
  }

  \caption{\textsc{Recycling Space-Saving-Sets($s$)}\label{alg:rsss}}
\end{algorithm}

In practice, the recycling variant does slightly better when labels share a lot
of items as it less likely to over-count in this case. Both variants do poorly on
sets whose ranks are close to the size parameter $s$ of the sketch. Thus on
particularly challenging data sets, for $s=1000$, only the top 100 might have
accurate estimates, while lower ranked sets have error closer to $m/s$. The poor
estimates are caused by the large amount of churn for the smaller sets in the
sketch (as we always replace the minimum counter), and the estimates can quickly
become over-inflated.

\subsection{Sampling the Input}

The sampling-based solutions for the heavy distinct hitter problem mentioned in
Section~\ref{subsec:hdh_previous_work} avoid over-estimating the cardinality of
small sets as the probability of sampling them is low. The idea of combining
sketching with sampling has appeared in other domains recently \cite{CPW-2020},
\cite{SFB-2022}, and the aforementioned idwsHH~\cite{FABCS-2017} by Feibish et
al. recently introduced the idea for heavy distinct hitters.


We modify Algorithm~\ref{alg:rsss} to sample the input. Let $h$ be a hash
function that uniformly maps to the open unit interval $U(0,1)$. For an input
$(\ell, x)$ we use $1/h(x)$ as an estimate for $d_{\ell}$ (the number of
distinct items with label $\ell$). Note that if we hash $d_{\ell}$ different
items the expected maximum value of $1/h(x)$ is $d_{\ell}+1$, as the minimum of
$n$ uniform $(0,1)$ variables is Beta$(1,n)$ distributed with expectation
$1/(n+1)$.

For any item $x$ and $\alpha > 1$,
$\Pr[1/h(x) < \alpha] = 1 - 1/\alpha$. Thus for labels corresponding to small
sets the chances of getting into the sketch shrink as the size of the minimum
count distinct sketch starts to grow.

\begin{algorithm}
  \Fn{\textsc{Insert}($\ell$, x)}{
    \If{$\ell\in S$;}{
      $S[\ell]$.\textsc{Insert}$(x)$\;
    } \Else {
      \If{$|S| < s$;}{
        $S.\textsc{Add}(\ell,\textsc{NewSketch()})$\;
        $S[\ell]$.\textsc{Insert}$(x)$\;
      } \Else {
        $y \leftarrow \arg \min_{j\in S} S[j].$\textsc{Distinct}$()$\;
        $s_y \leftarrow S[y].$\textsc{Distinct}$()$\;
        \If{$1/h(x) > s_y$;}{
          $S.\textsc{Add}(\ell,S[y])$\;
          $S.\textsc{Remove}(y)$\;
          $S[\ell]$.\textsc{Insert}$(x)$\;
        }
      }
    }
  }
  \caption{\textsc{Sampling Space-Saving-Sets($s$)}\label{alg:sampling}}
\end{algorithm}

We show an analogous lemma to Lemma~\ref{lem:min-props} for
Algorithm~\ref{alg:sampling}.

\begin{lemma}\label{lem:min-props-2}
  After $m$ insertions, let $d_\ell$ be the number of distinct elements $x_i$
  with label $\ell$.  Let $\alpha := \min_{j\in S} S[j].\textsc{Distinct}()$,
  and let $\abs{\{\ell : d_\ell > 0\}} > s$. Then for
  Algorithm~\ref{alg:sampling}:

  \begin{enumerate}
  \item $\alpha$ never decreases as entries are inserted;
  \item $\alpha \leq (m/s)(1+\eps)$;
  \item $\textsc{Query}(\ell) \leq d_\ell(1+\eps) + \alpha$; and
  \item If $\ell\not\in S$, then $d_\ell \leq \alpha\ln(1/\delta_r)$.
  \end{enumerate}
  The second and third properties hold with probability at least $1-\delta_c$ by
  the strong-tracking property of the count distinct sketch with accuracy
  guarantee $\eps$, while the fourth holds with probability at least $1-
  \delta_c - \delta_r$, for $\delta_r > 0$,
\end{lemma}
\begin{proof}
  The proofs of the first three properties remain unchanged
  from the proofs for Lemma~\ref{lem:min-props}.

  For the last property, consider the final time the counter for $\ell$ gets
  removed from $S$. At the time of removal, its value is the minimum among all
  the counters, and it can be upper-bounded by $\alpha$. Our main concern is
  with the number of items for $\ell$ that are sampled away after the sketch
  stops updating. If there are $r$ such remaining items, then the probability
  that they never enter the sketch is at most $(1-1/\alpha)^{r} <
  e^{-r/\alpha} = \delta_r$ for $r = \alpha\ln(1/\delta_r) $.
\end{proof}

\begin{theorem}
  For any label $\ell$, let $d_\ell$ be the number of distinct elements $x_i$
  with label $\ell$ in the stream. After $m$ insertions, to
  Algorithm~\ref{alg:sampling} with parameter $s$,
  \[\abs{\textsc{Query}(\ell) - d_\ell} \leq
  \Paren{\frac{m}{s}}(1+\eps) + m \max\Paren{\frac{(\ln(1/\delta_r)-1)(1+\eps)}{s}, \eps},\]
  with
  probability at least $1-\delta_c-\delta_r$ for $\delta_c,\delta_r > 0$.
\end{theorem}
\begin{proof}
  If $\ell\not \in S$, by properties 2 and 4 from Lemma~\ref{lem:min-props-2}, we
  have that the error is at most:
  \begin{eqnarray*}
  |\alpha - d_\ell| &
  \leq & \max(\alpha, d_\ell)
  \leq \alpha \ln(1/\delta_r)\\
  & \leq &
  \Paren{\frac{m}{s}}\Paren{1+\eps}\ln(1/\delta_r)\\
  & = &
  \Paren{\frac{m}{s}}\Paren{1+\eps} + m \Paren{\frac{(\ln(1/\delta_r)-1)(1+\eps)}{s}}.
  \end{eqnarray*}

  If $\ell\in S$, by properties 2 and 3 from Lemma~\ref{lem:min-props-2}, we have
  that the error is at most
  \[
  d_\ell \eps + \alpha
  \leq
  d_\ell \eps + \Paren{\frac{m}{s}}\Paren{1+ \eps}
  \leq
  m \eps + \Paren{\frac{m}{s}}\Paren{1+ \eps},
  \]
  where the last inequality comes from bounding $d_\ell \leq m$.
\end{proof}

\subsection{Practical Implementation}

Our final variant, the Practical Sampling Space-Saving Set sketch is the
implementation of Algorithm~\ref{alg:sampling} that we provide in our
open-source code and used in our experiments. Instead of calculating the minimum
size counter at every step, we only calculate the value when an item is
sampled. In our implementation we use HyperLogLog, a monotonic count distinct
sketch commonly used in practice with many highly optimized library
implementations~\cite{HNH-2013}, as the count distinct sketch.

\begin{algorithm}
  \Init{(s)}{
    $S \leftarrow $ empty associative array of strong-tracking,
    monotonic count distinct sketches with \textsc{NewSketch}, \textsc{Insert},
    \textsc{Distinct} methods\;

    $\theta \leftarrow 0.0$\;
  }

  \Fn{\textsc{Query}($\ell$)}{
    \If{$\ell \in S$} {
      \Return $S[\ell].\textsc{Distinct()}$\;
    } \Else {
      \Return $\min_{j\in S} S[j].\textsc{Distinct}()$
    }
  }

  \Fn{\textsc{Insert}($\ell$, x)}{
    \If{$\ell\in S$;}{
      $S[\ell]$.\textsc{Insert}$(x)$\;
    } \Else {
      \If{$|S| < s$;}{
        $S.\textsc{Add}(\ell,\textsc{NewSketch()})$\;
        $S[\ell]$.\textsc{Insert}$(x)$\;
      } \Else {
        \If{$1/h(x) > \theta$;}{
          $y \leftarrow \arg \min_{j\in S} S[j].$\textsc{Distinct}$()$\;
          $s_y \leftarrow S[y].$\textsc{Distinct}$()$\;
          $\theta = s_y$\;
          \If{$1/h(x) > s_y$;}{
            $S.\textsc{Add}(\ell,S[y])$\;
            $S.\textsc{Remove}(y)$\;
            $S[\ell]$.\textsc{Insert}$(x)$\;
          }
        }
      }
    }
  }
  \caption{\textsc{Practical Sampling Space-Saving-Sets($s$)}\label{alg:pssss}}
\end{algorithm}

Algorithm~\ref{alg:pssss} caches the size of the minimum counter into $\theta$,
so that we do not have to recalculate the size of the minimum counter on every
insert. Importantly, note that the caching does not affect the construction of
the sketch at all, and thus the outputs of Algorithms~\ref{alg:sampling}
and~\ref{alg:pssss} are equivalent.

\subsection{Mergeability}

Given two data streams, their corresponding Sampling Space-Saving Set sketches
can trivially be merged by merging corresponding count distinct sketches, then
only keeping the top $s$. We show the merge method for Algorithm~\ref{alg:pssss}
in Algorithm~\ref{alg:merge}.  The merge method for Algorithms~\ref{alg:osss},
\ref{alg:rsss}, and \ref{alg:sampling}, are all similar.

Merging is useful for both distributed stream processing, as well as aggregating
statistics over time---sketches for several consecutive time spans can be merged
to generate a sketch for the longer aggregate window.

\begin{algorithm}
  \Fn{\textsc{Merge}($other$)}{
    \For{ $\ell \in other.S$} {
      \If {$\ell \in S$} {
        $S[\ell]$.\textsc{Merge}($other.S[\ell]$)\;
      } \Else {
        $S.\textsc{Add}(\ell,other.S[\ell])$\;
      }
    }

    \tcc{sort by cardinality and take top $s$}
    \textsc{SortByValueDescending}$(S)$\;
    \While{ $|S| > s$ }{
      $S.\textsc{Pop}()$ \;
    }
    $\theta \leftarrow \min_{j\in S} S[j].\textsc{Distinct}()$\;
  }
  \caption{\textsc{Merging for Sampling Space-Saving-Sets($s$)}}
  \label{alg:merge}
\end{algorithm}

\section{Experiments}\label{sec:exp}

We evaluate Sampling Space-Saving Sets (SSSS) on several real-world and
synthetic data sets to demonstrate its state-of-the-art performance along
several axes.

\subsection{Metrics}
\label{subsec:metrics}

Previous experimental work evaluates the accuracy of heavy distinct hitter
sketches by comparing the output of a sketch to the actual set cardinalities
\emph{only} over the heavy sets, where ``heavy'' is determined by an arbitrary
threshold. The following example shows why such metrics are inherently flawed.

Let $d_i$ be the the number of distinct items with label $\ell_i$.  Consider the
simple example where the labels are in sorted order and, $d_1 = 100, d_2 =
100,..., d_{10} = 100, d_{11} = 10, ..., d_{20} = 10, d_{21} = 1, ..., d_{n}
=1$.  A sketch that returns the true values for the first 10 labels (i.e.,
$\tilde{d}_i = d_i, i \in \{1, ..., 10\}$), but also returns $\tilde{d}_i =
1000$ for $i \in \{11, ...., n\}$, will be considered to be perfectly accurate
under a metric that only evaluates accuracy over the heavy sets. If the sketch
is invertible, and one asks the sketch for the top 10 labels, it will respond
only with labels that are in fact \emph{outside} of the top 10.

To address this problem some papers~\cite{THL-2020} report precision and recall,
viewing the identification of heavy sets as a classification problem, but these
metrics are inherently flawed as well. We consider again the example above. If
the thresholds are set so that the top 20 are in the heavy set, a sketch
returning $\ell_1\through\ell_{10}, \ell_{21}\through\ell_{30}$ with zero error
but estimates of $0$ for all other labels will have exactly the same precision
and recall metrics as a sketch returning $\ell_{11}\through\ell_{30}$ with zero
error but estimates of $0$ for all other labels. Given the choice of the two, we
would much rather have the top 10 than the second 10, and we need an error
metric that can distinguish between the two cases.

To address both problems, we will report continuous error metrics not only
with respect to the actual heavy sets, but also with respect to the sketch's
self-reported heavy sets.

Some previous experimental works~\cite{Ting-2019} use relative error, $(d_i -
\tilde{d}_i)/d_i$, to show that the sketch does well in approximating
cardinalities across the board. If one evaluates the top 20 for the previous
example for a sketch that gets the first 10 labels exactly right, but outputs 0
for all the other labels, the relative accuracy will be just as bad as a sketch
that outputs 0 for the top 10 labels and outputs 10 for all the rest. In
practice, one would be much happier with the output of the first sketch as the
most common use cases for heavy distinct hitter sketches are in fact to better
approximate the heavier sets. For this reason, we will emphasize absolute error
$\abs{d_i - \tilde{d}_i}$ in the results, as it better captures the error over
the heavy items. We can further emphasize the heavy items by looking at the
squared error, though our guarantees are in terms of just the absolute
error.

As we will be running over multiple data sets with very different
characteristics, we will be normalizing our error metrics so that a perfect
sketch has an error of 0.0 and 1.0 corresponds to reporting 0 as the estimate
for every set cardinality. (Errors higher than 1.0 indicate that the sketch is
doing worse than the all-zero estimator.) Similarly, as the different data sets
have very different distributions, instead of specifying a ``heavy'' set using
an arbitrary threshold relative to a characteristic of the data set itself, we
will report the error over the top $k$ for $k = 10, 100, 1000$. Note that
considering the top $k$ is related to a threshold of $m/k$ (as there can be at
most $k$ sets with cardinality $m/k$). In other words, if we wish to measure
sets that comprise over 1\% of the stream, we would want to consider the top
100.

We will compare the sketches when they have the same memory constraints.  We
constrain the memory used by each sketch by setting its parameters to be as
large as possible while remaining under the memory limit. We will show the error
metrics over the top 10, 100, 1000 sets. We will notate the actual top $k$ sets
as $T_k$, and the top $k$ according to the sketch as $S_k$.

The precise error metrics we use will be Normalized Absolute Error throughout:
\[NAE(S) = \sum_{i\in S} \abs{d_i - \tilde{d}_i} / \sum_{i\in S} d_i\]
(where $S$ is either $T_k$ or $S_k$).

We include relative and squared error for completeness in our experiments, and
show that SSSS is superior regardless in the full version of this paper:
\begin{itemize}
\item Normalized Root Squared Error:
  \[NRSE(S) = \sqrt{\sum_{i\in S} \paren{d_i - \tilde{d}_i}^2 / \sum_{i\in S} d_i^2}\]
\item Relative Mean Absolute Error:
  \[RMAE(S) = |S|\inv\sum_{i\in S} \abs{d_i - \tilde{d}_i}/|d_i|\]
\item Relative Root Mean Square Error:
  \[RRMSE(S) = |S|\inv\sqrt{\sum_{i\in S} \paren{d_i - \tilde{d}_i}^2/|d_i|}\]
\end{itemize}

To combine the errors over the actual heavy sets and the sketch's heavy sets
into a single metric, we use the widely used quadratic mean ($\sqrt{(a^2 + b^2)/2}$):

\[Q_k: = \sqrt{(NAE(S_k)^2 +   NAE(T_k)^2)/2}.\]

The quadratic mean emphasizes the larger of the two numbers, and thus we bias
our metric towards the worse error. This is similar to the F1-score (the
harmonic mean) used to combine precision and recall, which emphasizes the lower
of the two numbers as higher values are better for precision and recall.

Returning to the first scenario we considered earlier in this section, the
algorithm that returns an estimate of 1000 for everything outside of the top 10
would have $NAE(T_{10}) = 0$, $NAE(S_{10}) = 999$, and thus $Q_{10} = 999/\sqrt{2}$.
For the second scenario, the two sketches would have
$NAE(S_{20}) = 0$, but the first sketch would have $NAE(T_{20})=1/11$ and
$Q_{20} \approx 0.06$ while the second sketch would have $NAE(T_{20})=10/11$ and
$Q_{20} \approx 0.64$.

\subsection{Configuration}
\label{subsec:configuration}

We compare Sampling Space-Saving Sets (SSSS) with the state-of-the-art heavy
distinct hitter sketch SpreadSketch~\cite{THL-2020} and the state-of-the-art
many count distinct sketch Count-HLL~\cite{Ting-2019}.  All three are
mergeable. SSSS and SpreadSketch are invertible, while Count-HLL in its original
form is not. Our results in Section~\ref{subsec:results} will be based on the
data-independent configurations described below. We do not provide a comparison
with idwsHH~\cite{FABCS-2017} in this section even though it is the algorithm
most similar to ours as it is not mergeable, and as noted in
Section~\ref{subsec:hdh_previous_work}, its accuracy is several orders of
magnitude worse than all the other algorithms.

Recall that SpreadSketch is a $d$ by $w$ array of count distinct sketches
(Section~\ref{subsec:hdh_previous_work}). It has 3 parameters: the depth
(which is set to 4 in their experiments), the width (which they vary to match
the memory limit), and the number of registers used by the count distinct sketch
(which they set to 438). Our experiments showed that SpreadSketch benefits
greatly from having as large a width as possible, so we use an even lower size
for the count distinct sketch of 64, and only vary the width.

Count-HLL also has depth and width parameters. In their experiments, they set
$d$ to be 512, and vary the width. We find 512 to be a reasonable setting as
well. As it is not invertible, we augment that algorithm using the technique
used by SpreadSketch where they keep $d \cdot w$ labels. When a particular cell
is updated, the label is updated if a rough estimate of the cardinality of that
label is greater than the rough estimate previously recorded for the label
corresponding to that cell. When asked for the heavy distinct hitters, the
sketch estimates cardinalities for all the labels it knows about, and outputs
the highest ones.

Finally, for SSSS our recommendation would be to set the count distinct sketch
size at 1024. Tuning SSSS is relatively straight-forward as the number of labels
kept by the sketch is the same as the number of count distinct sketches in the
sketch. As we shall see later in the experiments, in practice, SSSS's accuracy
quickly becomes the accuracy of the underlying count distinct sketch.
SpreadSketch and Count-HLL are more difficult to tune as a large part of their
memory budget is spent on holding an array of labels whose dimensions changes
along with the parameters being tuned for accuracy.

\subsection{Data}

We consider the performance of the three algorithms on three real data sets and
one synthetic data set. The first data set is the CAIDA UCSD Witty
Worm~\cite{CAIDA-WITTY} data set, where we use the first four hours of network
trace data. The labels here are the source IPs and the items are the destination
IPs. The task is to identify the high-cardinality sources. The next two data
sets (PubMed and KASANDR) are from the UCI Machine Learning
Repository~\cite{DuaGraff-2019}.  The Bag of Words PubMed abstracts data set is
a set of documents. The labels here are the words, the items are the documents,
and the task is to find the words appearing in the most documents. The KASANDR
data set~\cite{SLAVB-2017} is an ad impression data set where the labels are ad
offer ids and the items are user ids. The task is to find the offers that were
shown to the most unique users. The synthetic data set is created to be a
particularly difficult case for heavy distinct hitter sketches. It has labels
drawn according to the $Zipf(N;s)$ distribution for $N = 10^8$ and $s=0.2$, and
items are randomly chosen. We choose the exponent parameter $s$ to be much less
than 1.0 so that the tail falls very gradually.

Table~\ref{table:datasets} shows the basic statistics for each data set. An
\emph{entry} is a single (possibly non-unique) label-item pair in the data
stream.

\begin{table}[!ht]
  \footnotesize
  \centering
  \begin{tabular}{|c|c c c c|}
    \hline
                           & Witty        & PubMed       & KASANDR     & Zipf        \\
    \hline
    entries                & 123,931,016  & 483,450,157  & 15,844,717  & 100,000,000 \\
    entries for top 100    & 26\%         & 12\%         & 5\%         & 0.4\%       \\
    entries for top 1000   & 88\%         & 45\%         & 14\%        & 3\%         \\
    unique entries         & 99\%         & 100\%        & 52\%        & 100\%       \\
    unique labels          & 16,821       & 141,043      & 2,158,859   & 100,000     \\
    unique items           & 14,962,090   & 8,200,000    & 291,485     & 100,000,000 \\
    \hline
    avg label set size     & 7322.6       & 3,427.7      & 3.8         & 1000        \\
    p50 label set size     & 84           & 191          & 2           & 922         \\
    p90 label set size     & 6,311        & 3,278        & 7           & 1271        \\
    p99 label set size     & 195,592      & 66,354       & 33          & 2010        \\
    max label set size     & 502,142      & 2,323,263    & 4080        & 11,610      \\
    \hline
  \end{tabular}
  \caption{Descriptive statistics of the four data sets.}
  \label{table:datasets}
\end{table}

We can see that the KASANDR data set is an order of magnitude smaller than the
others, but it has an order of magnitude more labels than the PubMed data set,
which in turn has an order of magnitude more labels than the Witty data set. All
three real data sets have set size distributions that tail off sharply. To
contrast with the real data sets, we set the parameters of the Zipf data set so
that the tail is much fatter, and the average and median set sizes are within a
factor of 11 of the maximum-size set.

Finally, we also consider synthetic data sets that test performance when there is
a high amount of overlap among the sets with different labels. (These data sets
come from the paper for Count-HLL~\cite{Ting-2019}.)  The data is drawn from a
universe of $10^6$ items with $10^5$ of those items fixed as the common items
over which the sets overlap. We draw $k = 10^5$ or $10^6$ sets of size
1000 from the common items and 1000 sets of size $n$ from the full
universe where we vary $n$ from 20,000 to 500,000.

\subsection{Results}\label{subsec:results}

\begin{figure*}[ht]
  \centering
\def\axisdefaultheight{160pt}

\begin{tikzpicture}
  \begin{semilogyaxis}[
      footnotesize,
      title={\footnotesize{Witty}},
      cycle list name=dd,
      legend columns=-1,
      legend entries={SSSS, Count-HLL, Spread},
      legend to name={leg:a},
      ylabel={$NAE(Q_{10})$},
      yticklabel style={/pgf/number format/fixed},
      ytick={0.1, 0.01},
      yticklabels={$0.1$, $0.01$},
      ymin={0.01},
      ymax={0.5},
    ]
    \addplot table [x=mb, y=ssss_10] {accuracy_witty.dat};
    \addplot table [x=mb, y=chll_10] {accuracy_witty.dat};
    \addplot table [x=mb, y=spread_10] {accuracy_witty.dat};
  \end{semilogyaxis}
\end{tikzpicture}
\begin{tikzpicture}
  \begin{semilogyaxis}[
      footnotesize,
      title={\footnotesize{PubMed}},
      cycle list name=dd,
      yticklabel style={/pgf/number format/fixed},
      ytick={10.0, 1.0, 0.1, 0.01},
      yticklabels={$10.0$, $1.0$, $0.1$, $0.01$},
      ymin={0.01},
      ymax={1.0},
    ]
    \addplot table [x=mb, y=ssss_10] {accuracy_pubmed.dat};
    \addplot table [x=mb, y=chll_10] {accuracy_pubmed.dat};
    \addplot table [x=mb, y=spread_10] {accuracy_pubmed.dat};
  \end{semilogyaxis}
\end{tikzpicture}
\begin{tikzpicture}
  \begin{semilogyaxis}[
      footnotesize,
      title={\footnotesize{KASANDR}},
      cycle list name=dd,
      ytick={10.0, 1.0, 0.1, 0.01},
      yticklabels={$10.0$, $1.0$, $0.1$, $0.01$},
      ymin={0.01},
      ymax={64.0},
    ]
    \addplot table [x=mb, y=ssss_10] {accuracy_kasandr.dat};
    \addplot table [x=mb, y=chll_10] {accuracy_kasandr.dat};
    \addplot table [x=mb, y=spread_10] {accuracy_kasandr.dat};
    \draw [ gray, thin, dashed] (0,1.0) -- (6, 1.0);
  \end{semilogyaxis}
\end{tikzpicture}
\begin{tikzpicture}
  \begin{semilogyaxis}[
      footnotesize,
      title={\footnotesize{Zipf}},
      cycle list name=dd,
      ytick={10.0, 1.0, 0.1, 0.01},
      yticklabels={$10.0$, $1.0$, $0.1$, $0.01$},
      ymin={0.01},
      ymax={64.0},
    ]
    \addplot table [x=mb, y=ssss_100] {accuracy_zipf.dat};
    \addplot table [x=mb, y=chll_100] {accuracy_zipf.dat};
    \addplot table [x=mb, y=spread_100] {accuracy_zipf.dat};
    \draw [ gray, thin, dashed] (0,1.0) -- (6, 1.0);
  \end{semilogyaxis}
\end{tikzpicture}
\\

\begin{tikzpicture}
  \begin{semilogyaxis}[
      footnotesize,
      cycle list name=dd,
      ylabel={$NAE(Q_{100})$},
      yticklabel style={/pgf/number format/fixed},
      ytick={0.1, 0.01},
      yticklabels={$0.1$, $0.01$},
      ymin={0.01},
      ymax={0.5},
    ]
    \addplot table [x=mb, y=ssss_100] {accuracy_witty.dat};
    \addplot table [x=mb, y=chll_100] {accuracy_witty.dat};
    \addplot table [x=mb, y=spread_100] {accuracy_witty.dat};
  \end{semilogyaxis}
\end{tikzpicture}
\begin{tikzpicture}
  \begin{semilogyaxis}[
      footnotesize,
      cycle list name=dd,
      yticklabel style={/pgf/number format/fixed},
      ytick={10.0, 1.0, 0.1, 0.01},
      yticklabels={$10.0$, $1.0$, $0.1$, $0.01$},
      ymin={0.01},
      ymax={1.0},
    ]
    \addplot table [x=mb, y=ssss_100] {accuracy_pubmed.dat};
    \addplot table [x=mb, y=chll_100] {accuracy_pubmed.dat};
    \addplot table [x=mb, y=spread_100] {accuracy_pubmed.dat};
  \end{semilogyaxis}
\end{tikzpicture}
\begin{tikzpicture}
  \begin{semilogyaxis}[
      footnotesize,
      cycle list name=dd,
      ytick={10.0, 1.0, 0.1, 0.01},
      yticklabels={$10.0$, $1.0$, $0.1$, $0.01$},
      ymin={0.01},
      ymax={64.0},
    ]
    \addplot table [x=mb, y=ssss_100] {accuracy_kasandr.dat};
    \addplot table [x=mb, y=chll_100] {accuracy_kasandr.dat};
    \addplot table [x=mb, y=spread_100] {accuracy_kasandr.dat};
    \draw [ gray, thin, dashed] (0,1.0) -- (6, 1.0);
  \end{semilogyaxis}
\end{tikzpicture}
\begin{tikzpicture}
  \begin{semilogyaxis}[
      footnotesize,
      cycle list name=dd,
      ytick={10.0, 1.0, 0.1, 0.01},
      yticklabels={$10.0$, $1.0$, $0.1$, $0.01$},
      ymin={0.01},
      ymax={64.0},
    ]
    \addplot table [x=mb, y=ssss_100] {accuracy_zipf.dat};
    \addplot table [x=mb, y=chll_100] {accuracy_zipf.dat};
    \addplot table [x=mb, y=spread_100] {accuracy_zipf.dat};
    \draw [ gray, thin, dashed] (0,1.0) -- (6, 1.0);
  \end{semilogyaxis}
\end{tikzpicture}
\\

\begin{tikzpicture}
  \begin{semilogyaxis}[
      footnotesize,
      cycle list name=dd,
      xlabel={Memory (MiB)},
      ylabel={$NAE(Q_{1000})$},
      yticklabel style={/pgf/number format/fixed},
      ytick={0.1, 0.01},
      yticklabels={$0.1$, $0.01$},
      ymin={0.01},
      ymax={0.5},
    ]
    \addplot table [x=mb, y=ssss_1000] {accuracy_witty.dat};
    \addplot table [x=mb, y=chll_1000] {accuracy_witty.dat};
    \addplot table [x=mb, y=spread_1000] {accuracy_witty.dat};
  \end{semilogyaxis}
\end{tikzpicture}
\begin{tikzpicture}
  \begin{semilogyaxis}[
      footnotesize,
      cycle list name=dd,
      xlabel={Memory (MiB)},
      ytick={10.0, 1.0, 0.1, 0.01},
      yticklabels={$10.0$, $1.0$, $0.1$, $0.01$},
      ymin={0.01},
      ymax={1.0},
    ]
    \addplot table [x=mb, y=ssss_1000] {accuracy_pubmed.dat};
    \addplot table [x=mb, y=chll_1000] {accuracy_pubmed.dat};
    \addplot table [x=mb, y=spread_1000] {accuracy_pubmed.dat};
  \end{semilogyaxis}
\end{tikzpicture}
\begin{tikzpicture}
  \begin{semilogyaxis}[
      footnotesize,
      cycle list name=dd,
      xlabel={Memory (MiB)},
      ytick={10.0, 1.0, 0.1, 0.01},
      yticklabels={$10.0$, $1.0$, $0.1$, $0.01$},
      ymin={0.01},
      ymax={64.0},
    ]
    \addplot table [x=mb, y=ssss_1000] {accuracy_kasandr.dat};
    \addplot table [x=mb, y=chll_1000] {accuracy_kasandr.dat};
    \addplot table [x=mb, y=spread_1000] {accuracy_kasandr.dat};
    \draw [ gray, thin, dashed] (0,1.0) -- (6, 1.0);
  \end{semilogyaxis}
\end{tikzpicture}
\begin{tikzpicture}
  \begin{semilogyaxis}[
      footnotesize,
      cycle list name=dd,
      xlabel={Memory (MiB)},
      ytick={10.0, 1.0, 0.1, 0.01},
      yticklabels={$10.0$, $1.0$, $0.1$, $0.01$},
      ymin={0.01},
      ymax={64.0},
    ]
    \addplot table [x=mb, y=ssss_1000] {accuracy_zipf.dat};
    \addplot table [x=mb, y=chll_1000] {accuracy_zipf.dat};
    \addplot table [x=mb, y=spread_1000] {accuracy_zipf.dat};
    \draw [ gray, thin, dashed] (0,1.0) -- (6, 1.0);
  \end{semilogyaxis}
\end{tikzpicture}\\
\ref{leg:a}
\Description[accuracy graphs]{line graphs showing the accuracy of the sketches}
  \caption{Accuracy vs Memory Usage. $NAE(Q_{10}), NAE(Q_{100}), NAE(Q_{1000})$ as defined in
    Section~\ref{subsec:metrics} for SSSS, Count-HLL, and SpreadSketch. }
  \label{fig:graphs}
\end{figure*}
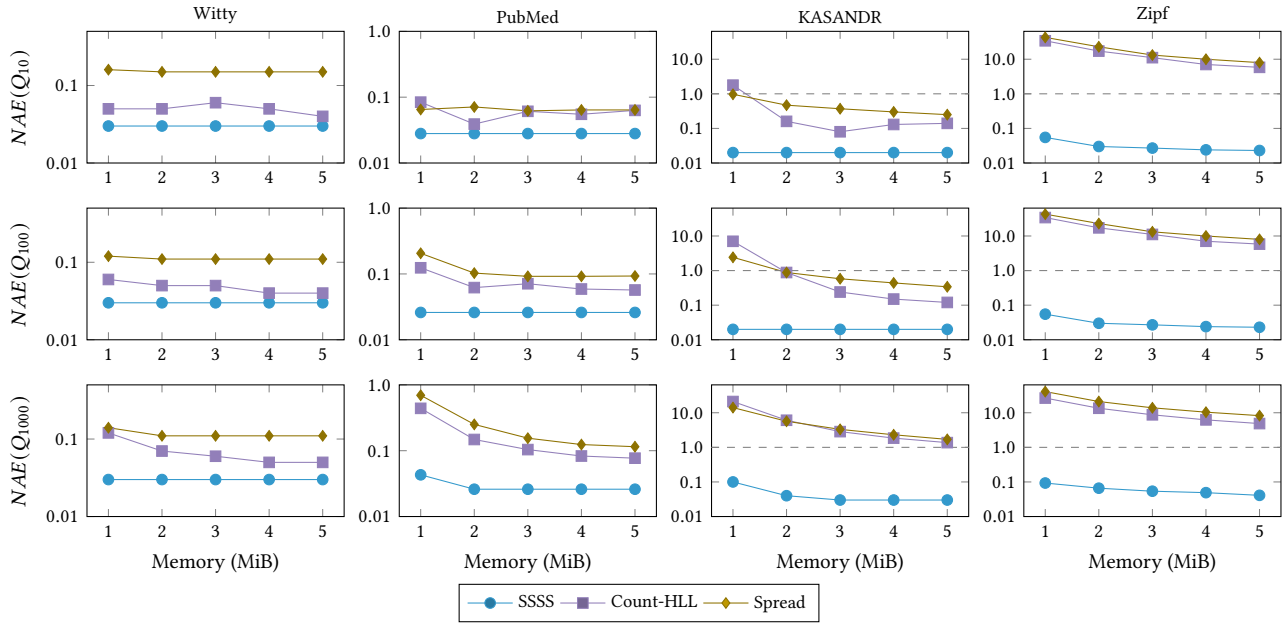

Sampling Space-Saving Sets can be seen to have superior accuracy
across the board over the state-of-the-art heavy distinct hitter and many count
distinct sketches in Figure~\ref{fig:graphs}. N.B., the accuracy graphs are
log-scaled on the $y$-axis to be able to reasonably present the errors of the
other sketches, which are much worse.

Witty, with the smallest overall number of labels and 88\% of its entries coming
from the top 1000 labels, is seen to be the easiest data set for all three
algorithms. Though even here SSSS with a consistent 0.03 NAE is superior to
CountHLL with 0.05-0.12 NAE and SpreadSketch with 0.08-0.13 NAE.

At the other extreme, KASANDR is the most difficult real data set as the number
of labels is the same magnitude as the number of entries, and only 14\% of the
entries come from the top 1000 labels. SpreadSketch only starts doing better
than the all-zero estimator (which would have an NAE of 1.0) when it has more
than 1.0 MiB to work with and even then only over the top 100 or top
10. Count-HLL also does worse than the all-zero estimator over the top 1000.

While SSSS only needs to hold a single label for each count distinct sketch,
Count-HLL and SpreadSketch need to hold $w$ and $d \cdot w$ times as many labels
respectively. This leaves less memory for count distinct sketches. Recall that
SpreadSketch is run with a redundancy of $d=4$. Thus, the effective number of
count distinct sketches is lower than that of SSSS, and the count distinct sketches
used by SpreadSketch only have 64 registers as opposed to the 1024 registers of
SSSS.

Overall, we see that SSSS quickly reaches the error inherent in its count distinct
sketch, which is low due to the relatively high number of registers. Count-HLL
performs remarkably well at lower memory sizes where it effectively has fewer
than 1000 count distinct sketches yet still manages to have fairly low NAE for the
Witty and PubMed data sets. For data sets with a lot more labels than its
capacity, Count-HLL starts to degrade. For data sets where SpreadSketch reaches
the error inherent in its count distinct sketch, it suffers from the low number of
registers in its count distinct sketch. For data sets with more labels than its
capacity, SpreadSketch quickly becomes noisy.

In the Zipf data set, the set sizes get gradually smaller so that there is not a
clear difference between ``signal'' sets and ``noise'' sets. We see that the
other sketches can not handle such a setting, and their accuracies are always
worse than the all-zero estimator. SSSS has no such problem.

To see how much of SSSS's superior performance is coming from only holding as
many labels as counters and thus having more space for both more and larger
counters, we look at the performance of the different sketches when they are all
set to have the same size (or ``width'') and the same count distinct sketch
size. Table~\ref{table:same_size} shows the performance for the three sketches
on the KASANDR data set for $s=w=2000$ and count distinct sketches of size
1024. Even here we see that SSSS has far superior accuracy while using 4 to 5
times less memory.

\begin{table}[!ht]
  \captionsetup{aboveskip=2pt, belowskip=6pt}
  \small
  \centering
  \begin{tabular}{|c|c c c|}
    \hline
                    & SSSS  & Count-HLL & Spread \\
    \hline
    $NAE(Q_{10})$    & 0.02  & 0.05  & 0.81 \\
    $NAE(Q_{100})$   & 0.02  & 0.07  &  1.6 \\
    $NAE(Q_{1000})$  & 0.04  & 0.59  &  10.0 \\
    Memory          & 2.2 MiB  & 11.7 MiB  &  8.6 MiB \\
    \hline
  \end{tabular}
  \caption{Accuracy and memory usage for SSSS, Count-HLL, and SpreadSketch
    when running on the KASANDR data set with the same size settings (log scale)}
  \label{table:same_size}
\end{table}

The next two tables explore how SSSS's behavior changes under different
settings. Table~\ref{table:vary_size} shows how the size parameter $s$ and
accuracy of SSSS vary when holding its memory constant at 3.0~MiB, while varying
the size of its underlying count distinct sketch (in this case HLL). We see that
in general it helps to increase the size of the distinct count sketch, but only
until the number of counters for SSSS does not drop too low---if we care about
the top 1000, it helps to have $s$ be at least 1000.

\begin{table}[!ht]
  \captionsetup{aboveskip=2pt, belowskip=6pt}
  \small
  \centering
  \begin{tabular}{|c|c c c c c c c|}
    \hline
    HLL size & 64    & 128   & 256  & 512  & 1024 & 2048 & 4096 \\
    \hline
    SSSS size $s$      & 18289 & 13329 & 8642 & 5073 & 2778 & 1459 & 748  \\
    $NAE(Q_{1000})$ & 0.096 & 0.083 & 0.049 & 0.038 & 0.034 & 0.047 & 0.096 \\
    \hline
  \end{tabular}
  \caption{The effect of varying the underlying count distinct sketch size for
    SSSS at 3.0 MiB when running on the KASANDR data set.}
  \label{table:vary_size}
\end{table}

Table~\ref{table:vary_algo} shows the difference in accuracy between
Algorithm~\ref{alg:osss} (SSS), Algorithm~\ref{alg:rsss} (RSSS), and
Algorithm~\ref{alg:sampling} (SSSS) for $s=2000$, showing the necessity of sampling
the input.

\begin{table}[!ht]
  \captionsetup{aboveskip=2pt, belowskip=6pt}
  \small
  \centering
  \begin{tabular}{|c|c c c|}
    \hline
     & SSS  & RSSS   & SSSS  \\
    \hline
    $NAE(Q_{10})$   & 1.2 & 0.68 & 0.02 \\
    $NAE(Q_{100})$  & 4.1 & 2.7  & 0.02 \\
    $NAE(Q_{1000})$ & 24.0  & 18.0   & 0.04 \\
    \hline
  \end{tabular}
  \caption{Accuracy when running SSS, RSSS, and SSSS at $s=2000$ on the KASANDR data set.}
  \label{table:vary_algo}
\end{table}

Next we explore the speed of the different sketches.
Table~\ref{table:throughput} shows the average throughput (number of entries
processed per millisecond) achieved by each sketch. SSSS and Count-HLL are
largely comparable, whereas SpreadSketch is the slowest of the three. (Note that
the throughput numbers for the real data sets include the time for data
I/O. More precise microbenchmarks are provided in our open-source prototype.)

\begin{table}[!ht]
  \captionsetup{aboveskip=2pt, belowskip=6pt}
  \small
  \centering
  \begin{tabular}{|c| c c c c|}
    \hline
    &  Witty  & PubMed & KASANDR & Zipf \\
    \hline
    SSSS       & 2970  &  3650  &  1600  & 21100 \\
    Count-HLL  & 2810  &  3570  &  1480  & 27900 \\
    Spread     & 2120  &  2470  &  1310  & 19700 \\
    \hline
  \end{tabular}
  \caption{Average throughput (entries processed per millisecond) for SSSS,
    Count-HLL, and SpreadSketch at 3 MiB.}
  \label{table:throughput}
\end{table}

\begin{figure*}[ht]
  \centering
\def\axisdefaultheight{160pt}

\begin{tikzpicture}
  \begin{semilogyaxis}[
      footnotesize,
      title={\footnotesize{Witty}},
      cycle list name=dd,
      legend columns=-1,
      legend entries={SSSS, Count-HLL, Spread},
      legend to name={leg:q},
      ylabel={seconds},
      xlabel={Memory (MiB)},
    ]
    \addplot table [x=mb, y=ssss] {query_witty.dat};
    \addplot table [x=mb, y=chll] {query_witty.dat};
    \addplot table [x=mb, y=spread] {query_witty.dat};
  \end{semilogyaxis}
\end{tikzpicture}
\begin{tikzpicture}
  \begin{semilogyaxis}[
      footnotesize,
      title={\footnotesize{PubMed}},
      cycle list name=dd,
      xlabel={Memory (MiB)},
    ]
    \addplot table [x=mb, y=ssss] {query_pubmed.dat};
    \addplot table [x=mb, y=chll] {query_pubmed.dat};
    \addplot table [x=mb, y=spread] {query_pubmed.dat};
  \end{semilogyaxis}
\end{tikzpicture}
\begin{tikzpicture}
  \begin{semilogyaxis}[
      footnotesize,
      title={\footnotesize{KASANDR}},
      cycle list name=dd,
      xlabel={Memory (MiB)},
    ]
    \addplot table [x=mb, y=ssss] {query_kasandr.dat};
    \addplot table [x=mb, y=chll] {query_kasandr.dat};
    \addplot table [x=mb, y=spread] {query_kasandr.dat};
  \end{semilogyaxis}
\end{tikzpicture}
\begin{tikzpicture}
  \begin{semilogyaxis}[
      footnotesize,
      title={\footnotesize{Zipf}},
      cycle list name=dd,
      xlabel={Memory (MiB)},
    ]
    \addplot table [x=mb, y=ssss] {query_zipf.dat};
    \addplot table [x=mb, y=chll] {query_zipf.dat};
    \addplot table [x=mb, y=spread] {query_zipf.dat};
  \end{semilogyaxis}
\end{tikzpicture}
\ref{leg:q}
\Description[query graphs]{line graphs showing query speed to top 1000}
  \caption{Top 1000 Query Duration (log scale)}
  \label{fig:query}
\end{figure*}
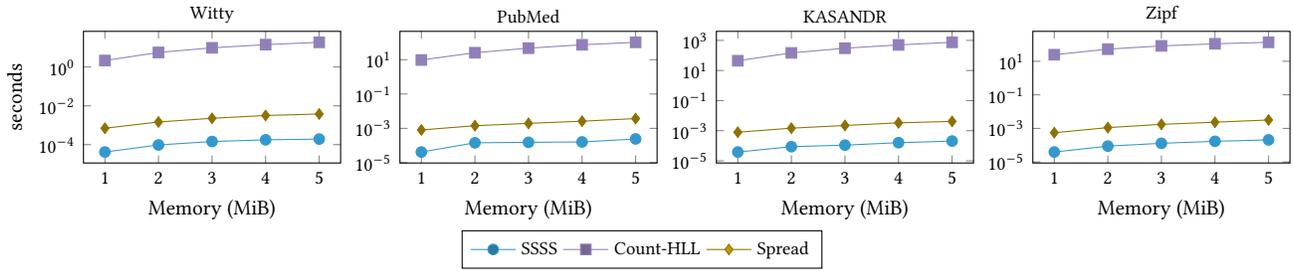

In addition to having the best accuracy relative to a given memory size, SSSS is
also far faster to query than the other two sketches. We compare the amount of
time it takes a sketch to produce the top 1000 labels along with estimates of
the cardinality of their corresponding sets (Figure~\ref{fig:query}). SSSS is
an order of magnitude faster than SpreadSketch, which in itself is several
orders of magnitude faster than Count-HLL. SSSS is faster as it keeps what it
thinks are the top labels and thus returning a top list is trivial. Both
SpreadSketch and Count-HLL have to estimate the cardinalities of all the labels
kept by the sketches and then return the top list from there. The
cardinality-estimation method of Count-HLL is much slower than that of
SpreadSketch as it involves estimating a maximum likelihood.

Next, we show how the different sketches perform when the data is segmented into
separate streams (124 for Witty, 484 for PubMed, 159 for KASANDR, and 100 for
Zipf), processed in parallel, and merged into a single sketch
(Figure~\ref{fig:merge}).  All three algorithms have worse accuracy than when
processing a single stream, and once again the KASANDR data set causes the most
difficulties among the real data sets. Count-HLL and SpreadSketch never achieve
$NAE(Q_{1000}) < 1.0$ even at 5.0~MiB, and do substantially worse at smaller
memory sizes. SSSS has $NAE(Q_{1000})$ ranging from 0.22 to 0.48, which is much
worse than the single stream version, but still an order of magnitude better
than the other sketches. The results for merging the synthetic Zipf datastreams
follow a similar pattern as all the sketches do worse, but only SSSS does better
than the all-zero estimator.

\begin{figure*}[ht]
  \centering
\def\axisdefaultheight{160pt}

\begin{tikzpicture}
  \begin{semilogyaxis}[
      footnotesize,
      title={\footnotesize{Witty}},
      cycle list name=dd,
      legend to name=leg:m,
      legend columns=-1,
      legend entries={SSSS, Count-HLL, Spread},
      ylabel={$NAE(Q_{10})$},
      yticklabel style={/pgf/number format/fixed},
      ytick={1.0, 0.1, 0.01},
      yticklabels={$1.0$, $0.1$, $0.01$},
      ymin={0.01},
      ymax={0.2},
    ]
    \addplot table [x=mb, y=ssss_10] {merge_witty.dat};
    \addplot table [x=mb, y=chll_10] {merge_witty.dat};
    \addplot table [x=mb, y=spread_10] {merge_witty.dat};
  \end{semilogyaxis}
\end{tikzpicture}
\begin{tikzpicture}
  \begin{semilogyaxis}[
      footnotesize,
      title={\footnotesize{PubMed}},
      cycle list name=dd,
      ytick={1.0, 0.1, 0.01},
      yticklabels={$1.0$, $0.1$, $0.01$},
      ymin={0.01},
      ymax={1.0},
      yticklabel style={/pgf/number format/fixed},
    ]
    \addplot table [x=mb, y=ssss_10] {merge_pubmed.dat};
    \addplot table [x=mb, y=chll_10] {merge_pubmed.dat};
    \addplot table [x=mb, y=spread_10] {merge_pubmed.dat};
  \end{semilogyaxis}
\end{tikzpicture}
\begin{tikzpicture}
  \begin{semilogyaxis}[
      footnotesize,
      title={\footnotesize{KASANDR}},
      cycle list name=dd,
      ytick={10.0, 1.0, 0.1, 0.01},
      yticklabels={$10.0$, $1.0$, $0.1$, $0.01$},
      ymin={0.01},
      ymax={64.0},
    ]
    \addplot table [x=mb, y=ssss_10] {merge_kasandr.dat};
    \addplot table [x=mb, y=chll_10] {merge_kasandr.dat};
    \addplot table [x=mb, y=spread_10] {merge_kasandr.dat};
    \draw [ gray, thin, dashed] (0,1.0) -- (6, 1.0);
    \end{semilogyaxis}
\end{tikzpicture}
\begin{tikzpicture}
  \begin{semilogyaxis}[
      footnotesize,
      title={\footnotesize{Zipf}},
      cycle list name=dd,
      ytick={100, 10.0, 1.0, 0.1, 0.01},
      yticklabels={$100$, $10$, $1.0$, $0.1$, $0.01$},
      ymin={0.01},
      ymax={128},
    ]
    \addplot table [x=mb, y=ssss_10] {merge_zipf.dat};
    \addplot table [x=mb, y=chll_10] {merge_zipf.dat};
    \addplot table [x=mb, y=spread_10] {merge_zipf.dat};
    \draw [ gray, thin, dashed] (0,1.0) -- (6, 1.0);
    \end{semilogyaxis}
\end{tikzpicture}
\\

\begin{tikzpicture}
  \begin{semilogyaxis}[
      footnotesize,
      cycle list name=dd,
      ylabel={$NAE(Q_{100})$},
      yticklabel style={/pgf/number format/fixed},
      ytick={1.0, 0.1, 0.01},
      yticklabels={$1.0$, $0.1$, $0.01$},
      ymin={0.01},
      ymax={0.2},
    ]
    \addplot table [x=mb, y=ssss_100] {merge_witty.dat};
    \addplot table [x=mb, y=chll_100] {merge_witty.dat};
    \addplot table [x=mb, y=spread_100] {merge_witty.dat};
  \end{semilogyaxis}
\end{tikzpicture}
\begin{tikzpicture}
  \begin{semilogyaxis}[
      footnotesize,
      cycle list name=dd,
      ytick={1.0, 0.1, 0.01},
      yticklabels={$1.0$, $0.1$, $0.01$},
      ymin={0.01},
      ymax={1.0},
      yticklabel style={/pgf/number format/fixed},
    ]
    \addplot table [x=mb, y=ssss_100] {merge_pubmed.dat};
    \addplot table [x=mb, y=chll_100] {merge_pubmed.dat};
    \addplot table [x=mb, y=spread_100] {merge_pubmed.dat};
  \end{semilogyaxis}
\end{tikzpicture}
\begin{tikzpicture}
  \begin{semilogyaxis}[
      footnotesize,
      cycle list name=dd,
      ytick={10.0, 1.0, 0.1, 0.01},
      yticklabels={$10.0$, $1.0$, $0.1$, $0.01$},
      ymin={0.01},
      ymax={64.0},
    ]
    \addplot table [x=mb, y=ssss_100] {merge_kasandr.dat};
    \addplot table [x=mb, y=chll_100] {merge_kasandr.dat};
    \addplot table [x=mb, y=spread_100] {merge_kasandr.dat};
    \draw [ gray, thin, dashed] (0,1.0) -- (6, 1.0);
    \end{semilogyaxis}
\end{tikzpicture}
\begin{tikzpicture}
  \begin{semilogyaxis}[
      footnotesize,
      cycle list name=dd,
      ytick={100, 10.0, 1.0, 0.1, 0.01},
      yticklabels={$100$, $10$, $1.0$, $0.1$, $0.01$},
      ymin={0.01},
      ymax={128},
    ]
    \addplot table [x=mb, y=ssss_100] {merge_zipf.dat};
    \addplot table [x=mb, y=chll_100] {merge_zipf.dat};
    \addplot table [x=mb, y=spread_100] {merge_zipf.dat};
    \draw [ gray, thin, dashed] (0,1.0) -- (6, 1.0);
    \end{semilogyaxis}
\end{tikzpicture}
\\

\begin{tikzpicture}
  \begin{semilogyaxis}[
      footnotesize,
      cycle list name=dd,
      xlabel={Memory (MiB)},
      ylabel={$NAE(Q_{1000})$},
      yticklabel style={/pgf/number format/fixed},
      ytick={1.0, 0.1, 0.01},
      yticklabels={$1.0$, $0.1$, $0.01$},
      ymin={0.01},
      ymax={0.2},
    ]
    \addplot table [x=mb, y=ssss_1000] {merge_witty.dat};
    \addplot table [x=mb, y=chll_1000] {merge_witty.dat};
    \addplot table [x=mb, y=spread_1000] {merge_witty.dat};
  \end{semilogyaxis}
\end{tikzpicture}
\begin{tikzpicture}
  \begin{semilogyaxis}[
      footnotesize,
      cycle list name=dd,
      xlabel={Memory (MiB)},
      ytick={1.0, 0.1, 0.01},
      yticklabels={$1.0$, $0.1$, $0.01$},
      ymin={0.01},
      ymax={1.0},
    ]
    \addplot table [x=mb, y=ssss_1000] {merge_pubmed.dat};
    \addplot table [x=mb, y=chll_1000] {merge_pubmed.dat};
    \addplot table [x=mb, y=spread_1000] {merge_pubmed.dat};
  \end{semilogyaxis}
\end{tikzpicture}
\begin{tikzpicture}
  \begin{semilogyaxis}[
      footnotesize,
      cycle list name=dd,
      xlabel={Memory (MiB)},
      ytick={10.0, 1.0, 0.1, 0.01},
      yticklabels={$10.0$, $1.0$, $0.1$, $0.01$},
      ymin={0.01},
      ymax={64.0},
    ]
    \addplot table [x=mb, y=ssss_1000] {merge_kasandr.dat};
    \addplot table [x=mb, y=chll_1000] {merge_kasandr.dat};
    \addplot table [x=mb, y=spread_1000] {merge_kasandr.dat};
    \draw [ gray, thin, dashed] (0,1.0) -- (6, 1.0);
    \end{semilogyaxis}
\end{tikzpicture}
\begin{tikzpicture}
  \begin{semilogyaxis}[
      footnotesize,
      cycle list name=dd,
      xlabel={Memory (MiB)},
      ytick={100, 10.0, 1.0, 0.1, 0.01},
      yticklabels={$100$, $10$, $1.0$, $0.1$, $0.01$},
      ymin={0.01},
      ymax={128},
    ]
    \addplot table [x=mb, y=ssss_1000] {merge_zipf.dat};
    \addplot table [x=mb, y=chll_1000] {merge_zipf.dat};
    \addplot table [x=mb, y=spread_1000] {merge_zipf.dat};
    \draw [ gray, thin, dashed] (0,1.0) -- (6, 1.0);
    \end{semilogyaxis}
\end{tikzpicture}\\
\ref{leg:m}
\Description[merge graphs]{line graphs showing accuracy under merging}
  \caption{Accuracy under merging (log scale)}
  \label{fig:merge}
\end{figure*}
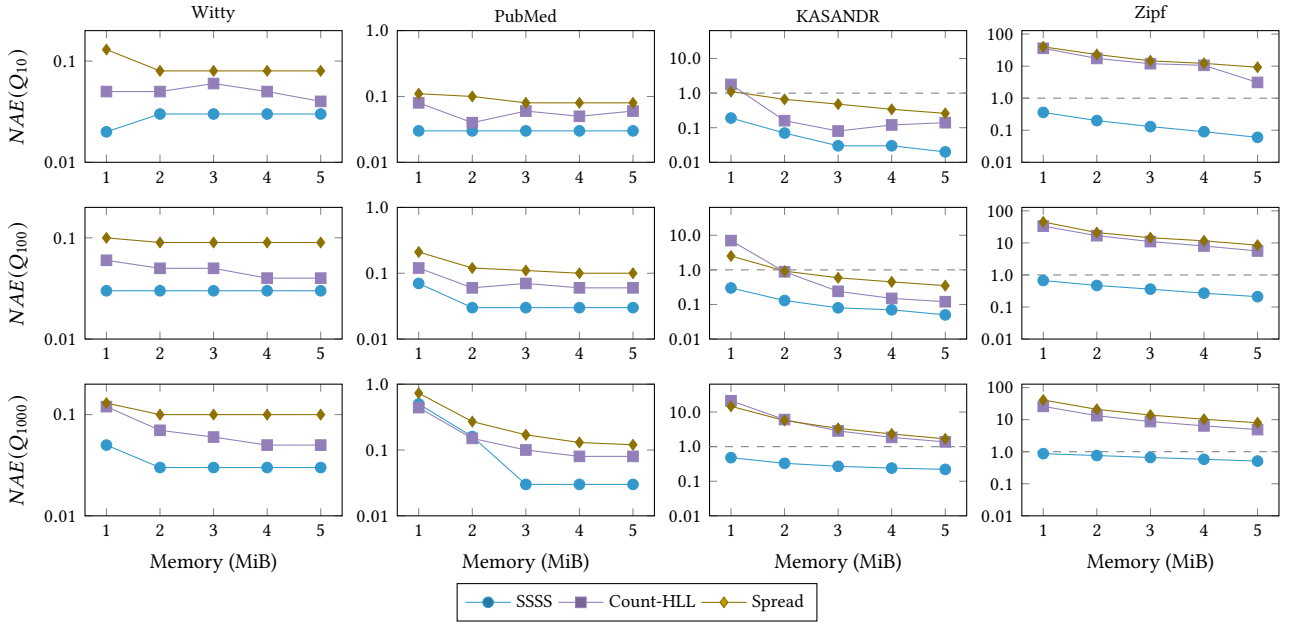

\begin{figure}[ht]
  \centering
\def\axisdefaultheight{160pt}

\begin{tikzpicture}
  \begin{loglogaxis}[
      footnotesize,
      title={\footnotesize{$k = 10^5$}},
      cycle list name=dd,
      legend columns=-1,
      legend entries={SSSS, Count-HLL, Spread},
      legend to name={leg:o},
      xlabel={$n$},
      xtick={20000, 50000,100000,200000,500000},
      xticklabels={$20K$,$50k$, $100k$, $200k$, $500k$},
      ylabel={$NAE(Q_{1000})$},
      ytick={10.0, 1.0, 0.1, 0.01},
      yticklabels={$10.0$, $1.0$, $0.1$, $0.01$},
      ymin={0.01},
      ymax={16.0},
      width={4.5cm},
    ]
    \addplot table [x=n_big, y=ssss_1000] {accuracy_overlap_10_5.dat};
    \addplot table [x=n_big, y=chll_1000] {accuracy_overlap_10_5.dat};
    \addplot table [x=n_big, y=spread_1000] {accuracy_overlap_10_5.dat};
    \draw [ gray, thin, dashed] (0,1.0) -- (700000, 1.0);
    \end{loglogaxis}
\end{tikzpicture}
\begin{tikzpicture}
  \begin{loglogaxis}[
      footnotesize,
      title={\footnotesize{$k = 10^6$}},
      cycle list name=dd,
      xlabel={$n$},
      xtick={20000, 50000,100000,200000,500000},
      xticklabels={$20K$,$50k$, $100k$, $200k$, $500k$},
      ytick={100.0, 10.0, 1.0, 0.1, 0.01},
      yticklabels={$100.0$, $10.0$, $1.0$, $0.1$, $0.01$},
      ymin={0.01},
      ymax={128.0},
      width={4.5cm},
    ]
    \addplot table [x=n_big, y=ssss_1000] {accuracy_overlap_10_6.dat};
    \addplot table [x=n_big, y=chll_1000] {accuracy_overlap_10_6.dat};
    \addplot table [x=n_big, y=spread_1000] {accuracy_overlap_10_6.dat};
    \draw [ gray, thin, dashed] (0,1.0) -- (700000, 1.0);
  \end{loglogaxis}
\end{tikzpicture}\\
\ref{leg:o}
\Description[query graphs]{line graphs showing the accuracy of the sketches over the overlap synthetic data set}
  \caption{Accuracy on synthetic sets with high overlap (log scale)}
  \label{fig:overlap}
\end{figure}
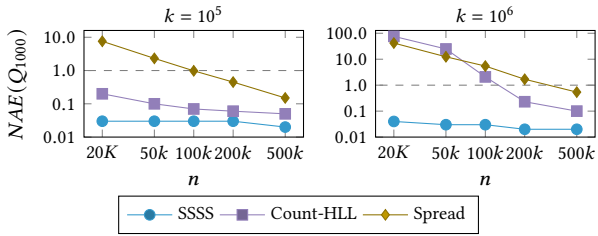

Finally, we show how the different algorithms perform on the synthetic overlap
data sets (Figure~\ref{fig:overlap}). Recall that there are exactly 1000 heavy
sets where we vary their sizes and $10^5$ or $10^6$ small sets of size 1000.  We
configure the sketches to be of size (or ``width'') exactly 1000 to match the
number of heavy sets. SSSS always performs best and reaches the error inherent
in its count distinct sketch. When there are $10^6$ small sets Count-HLL does worse
than the all-zero estimator until the heavy sets are 200 times the size of the
small sets. Count-HLL does much better when there are only $10^5$ small sets,
but it still performs 2 to 6 times worse than SSSS. SpreadSketch does worse than
the all-zero estimator until the heavy sets are 200 to 500 times the size of the
small sets. We also emphasize that SSSS achieves its outsize performance using
only 1.1 MiB, compared to 5.9 MiB for Count-HLL and 4.3 MiB for SpreadSketch.

\section{Conclusion}\label{sec:Conclusion}

We have shown that Sampling Space-Saving Sets (SSSS) is the only truly practical
sketch for the \emph{heavy distinct hitter} problem. The use of a Space-Saving
sketch allows SSSS to be easily invertible, unlike SpreadSketch or sketches for
the \emph{many distinct count} problem (e.g,  Count-HLL), which are only
invertible with the addition of an auxiliary structure for holding labels.
SSSS is the only option with practical query performance, being orders of
magnitude faster than its competitors.

In addition to its invertibility, SSSS is by far the most accurate sketch. We
showed its superior accuracy on both real data sets that exhibit the sharply
falling tails typical of large-scale data streams, but also on synthetic data
sets that have been constructed to have fat tails, and those that have been
constructed to have overlapping sets. We also show that SSSS is the most
accurate sketch when merging.

For future work, we believe that the worst-case bounds shown in this paper can
be greatly improved if we make some reasonable assumptions about the tail of the
input stream, or if we assume that the input stream is drawn i.i.d.

We have focused on insertion-only streams in this work as that corresponds to
high-throughput use-cases outlined in the introduction. There is also a rich
vein of work in the \emph{bounded-deletion} model \cite{JayaramWoodruff-2018,
  ZMWAA-2021, ZAAM-2022}. It would be interesting to see if the techniques
outlined here could be adapted to that model.

\bibliographystyle{ACM-Reference-Format}
\bibliography{heavy_sets}


\begin{thebibliography}{53}


\ifx \showCODEN    \undefined \def \showCODEN     #1{\unskip}     \fi
\ifx \showDOI      \undefined \def \showDOI       #1{#1}\fi
\ifx \showISBNx    \undefined \def \showISBNx     #1{\unskip}     \fi
\ifx \showISBNxiii \undefined \def \showISBNxiii  #1{\unskip}     \fi
\ifx \showISSN     \undefined \def \showISSN      #1{\unskip}     \fi
\ifx \showLCCN     \undefined \def \showLCCN      #1{\unskip}     \fi
\ifx \shownote     \undefined \def \shownote      #1{#1}          \fi
\ifx \showarticletitle \undefined \def \showarticletitle #1{#1}   \fi
\ifx \showURL      \undefined \def \showURL       {\relax}        \fi
\providecommand\bibfield[2]{#2}
\providecommand\bibinfo[2]{#2}
\providecommand\natexlab[1]{#1}
\providecommand\showeprint[2][]{arXiv:#2}

\bibitem[Agarwal et~al\mbox{.}(2012)]%
        {ACHPWY-2012}
\bibfield{author}{\bibinfo{person}{Pankaj~K. Agarwal}, \bibinfo{person}{Graham
  Cormode}, \bibinfo{person}{Zengfeng Huang}, \bibinfo{person}{Jeff Phillips},
  \bibinfo{person}{Zhewei Wei}, {and} \bibinfo{person}{Ke Yi}.}
  \bibinfo{year}{2012}\natexlab{}.
\newblock \showarticletitle{Mergeable Summaries}. In
  \bibinfo{booktitle}{\emph{Proceedings of the 31st ACM SIGMOD-SIGACT-SIGAI
  Symposium on Principles of Database Systems}} (Scottsdale, Arizona, USA)
  \emph{(\bibinfo{series}{PODS '12})}. \bibinfo{publisher}{ACM},
  \bibinfo{address}{New York, NY, USA}, \bibinfo{pages}{23--34}.
\newblock
\showISBNx{978-1-4503-1248-6}
\urldef\tempurl%
\url{https://doi.org/10.1145/2213556.2213562}
\showDOI{\tempurl}


\bibitem[Bar{-}Yossef et~al\mbox{.}(2002)]%
        {BJKST-2002}
\bibfield{author}{\bibinfo{person}{Ziv Bar{-}Yossef}, \bibinfo{person}{T.~S.
  Jayram}, \bibinfo{person}{Ravi Kumar}, \bibinfo{person}{D. Sivakumar}, {and}
  \bibinfo{person}{Luca Trevisan}.} \bibinfo{year}{2002}\natexlab{}.
\newblock \showarticletitle{Counting Distinct Elements in a Data Stream}. In
  \bibinfo{booktitle}{\emph{Randomization and Approximation Techniques, 6th
  International Workshop, {RANDOM} 2002, Cambridge, MA, USA, September 13-15,
  2002, Proceedings}} \emph{(\bibinfo{series}{Lecture Notes in Computer
  Science}, Vol.~\bibinfo{volume}{2483})},
  \bibfield{editor}{\bibinfo{person}{Jos{\'{e}} D.~P. Rolim} {and}
  \bibinfo{person}{Salil~P. Vadhan}} (Eds.). \bibinfo{publisher}{Springer},
  \bibinfo{address}{Berlin, Germany}, \bibinfo{pages}{1--10}.
\newblock
\urldef\tempurl%
\url{https://doi.org/10.1007/3-540-45726-7\_1}
\showDOI{\tempurl}


\bibitem[Beyer et~al\mbox{.}(2016)]%
        {BJPM-2016}
\bibfield{author}{\bibinfo{person}{Betsy Beyer}, \bibinfo{person}{Chris Jones},
  \bibinfo{person}{Jennifer Petoff}, {and} \bibinfo{person}{Niall~Richard
  Murphy}.} \bibinfo{year}{2016}\natexlab{}.
\newblock \bibinfo{booktitle}{\emph{Site Reliability Engineering: How Google
  Runs Production Systems}}.
\newblock \bibinfo{publisher}{"O'Reilly Media, Inc."},
  \bibinfo{address}{Sebastopol, CA, USA}.
\newblock


\bibitem[B\l{}asiok(2018)]%
        {Blasiok-2018}
\bibfield{author}{\bibinfo{person}{Jaros\l{}aw B\l{}asiok}.}
  \bibinfo{year}{2018}\natexlab{}.
\newblock \showarticletitle{Optimal streaming and tracking distinct elements
  with high probability}. In \bibinfo{booktitle}{\emph{Proceedings of the
  Twenty-Ninth Annual {ACM-SIAM} Symposium on Discrete Algorithms, {SODA} 2018,
  New Orleans, LA, USA, January 7-10, 2018}},
  \bibfield{editor}{\bibinfo{person}{Artur Czumaj}} (Ed.).
  \bibinfo{publisher}{{SIAM}}, \bibinfo{address}{Philadelphia, PA, USA},
  \bibinfo{pages}{2432--2448}.
\newblock
\urldef\tempurl%
\url{https://doi.org/10.1137/1.9781611975031.156}
\showDOI{\tempurl}


\bibitem[CAIDA(2004)]%
        {CAIDA-WITTY}
\bibfield{author}{\bibinfo{person}{CAIDA}.} \bibinfo{year}{2004}\natexlab{}.
\newblock \bibinfo{booktitle}{\emph{The {CAIDA} {UCSD} Dataset on the {W}itty
  {W}orm - March 19-24, 2004}}.
\newblock CAIDA UCSD.
\newblock


\bibitem[Cao et~al\mbox{.}(2009)]%
        {CJCBZ-2009}
\bibfield{author}{\bibinfo{person}{J. Cao}, \bibinfo{person}{Y. Jin},
  \bibinfo{person}{A. Chen}, \bibinfo{person}{T. Bu}, {and}
  \bibinfo{person}{Z.-L. Zhang}.} \bibinfo{year}{2009}\natexlab{}.
\newblock \showarticletitle{Identifying High Cardinality Internet Hosts}. In
  \bibinfo{booktitle}{\emph{IEEE INFOCOM 2009}}. \bibinfo{publisher}{{IEEE}},
  \bibinfo{address}{Piscataway, NJ, USA}, \bibinfo{pages}{810--818}.
\newblock
\urldef\tempurl%
\url{https://doi.org/10.1109/INFCOM.2009.5061990}
\showDOI{\tempurl}


\bibitem[Charikar et~al\mbox{.}(2002)]%
        {CCF-2002}
\bibfield{author}{\bibinfo{person}{Moses Charikar}, \bibinfo{person}{Kevin~C.
  Chen}, {and} \bibinfo{person}{Martin Farach{-}Colton}.}
  \bibinfo{year}{2002}\natexlab{}.
\newblock \showarticletitle{Finding Frequent Items in Data Streams}. In
  \bibinfo{booktitle}{\emph{Automata, Languages and Programming, 29th
  International Colloquium, {ICALP} 2002, Malaga, Spain, July 8-13, 2002,
  Proceedings}} \emph{(\bibinfo{series}{Lecture Notes in Computer Science},
  Vol.~\bibinfo{volume}{2380})}, \bibfield{editor}{\bibinfo{person}{Peter
  Widmayer}, \bibinfo{person}{Francisco~Triguero Ruiz},
  \bibinfo{person}{Rafael~Morales Bueno}, \bibinfo{person}{Matthew Hennessy},
  \bibinfo{person}{Stephan~J. Eidenbenz}, {and} \bibinfo{person}{Ricardo
  Conejo}} (Eds.). \bibinfo{publisher}{Springer}, \bibinfo{address}{Berlin,
  Germany}, \bibinfo{pages}{693--703}.
\newblock
\urldef\tempurl%
\url{https://doi.org/10.1007/3-540-45465-9\_59}
\showDOI{\tempurl}


\bibitem[Cohen et~al\mbox{.}(2020)]%
        {CPW-2020}
\bibfield{author}{\bibinfo{person}{Edith Cohen}, \bibinfo{person}{Rasmus Pagh},
  {and} \bibinfo{person}{David~P. Woodruff}.} \bibinfo{year}{2020}\natexlab{}.
\newblock \showarticletitle{{WOR} and \emph{p}'s: Sketches for
  {\(\mathscr{l}\)}\({}_{\mbox{p}}\)-Sampling Without Replacement}. In
  \bibinfo{booktitle}{\emph{Advances in Neural Information Processing Systems
  33: Annual Conference on Neural Information Processing Systems 2020, NeurIPS
  2020, December 6-12, 2020, virtual}}, \bibfield{editor}{\bibinfo{person}{Hugo
  Larochelle}, \bibinfo{person}{Marc'Aurelio Ranzato}, \bibinfo{person}{Raia
  Hadsell}, \bibinfo{person}{Maria{-}Florina Balcan}, {and}
  \bibinfo{person}{Hsuan{-}Tien Lin}} (Eds.). \bibinfo{publisher}{Curran
  Associates, Inc.}, \bibinfo{address}{Red Hook, NY},
  \bibinfo{pages}{21092--21104}.
\newblock
\urldef\tempurl%
\url{https://proceedings.neurips.cc/paper/2020/hash/f1507aba9fc82ffa7cc7373c58f8a613-Abstract.html}
\showURL{%
\tempurl}


\bibitem[Considine et~al\mbox{.}(2009)]%
        {CHLBK-2009}
\bibfield{author}{\bibinfo{person}{Jeffrey Considine}, \bibinfo{person}{Marios
  Hadjieleftheriou}, \bibinfo{person}{Feifei Li}, \bibinfo{person}{John~W.
  Byers}, {and} \bibinfo{person}{George Kollios}.}
  \bibinfo{year}{2009}\natexlab{}.
\newblock \showarticletitle{Robust approximate aggregation in sensor data
  management systems}.
\newblock \bibinfo{journal}{\emph{{ACM} Trans. Database Syst.}}
  \bibinfo{volume}{34}, \bibinfo{number}{1} (\bibinfo{year}{2009}),
  \bibinfo{pages}{6:1--6:35}.
\newblock
\urldef\tempurl%
\url{https://doi.org/10.1145/1508857.1508863}
\showDOI{\tempurl}


\bibitem[Cormode and Hadjieleftheriou(2010)]%
        {CormodeMuthukrishnan-2010}
\bibfield{author}{\bibinfo{person}{Graham Cormode} {and}
  \bibinfo{person}{Marios Hadjieleftheriou}.} \bibinfo{year}{2010}\natexlab{}.
\newblock \showarticletitle{Methods for Finding Frequent Items in Data
  Streams}.
\newblock \bibinfo{journal}{\emph{The VLDB Journal}} \bibinfo{volume}{19},
  \bibinfo{number}{1} (\bibinfo{date}{feb} \bibinfo{year}{2010}),
  \bibinfo{pages}{3–20}.
\newblock
\showISSN{1066-8888}
\urldef\tempurl%
\url{https://doi.org/10.1007/s00778-009-0172-z}
\showDOI{\tempurl}


\bibitem[Cormode and Muthukrishnan(2004)]%
        {CormodeMuthukrishnan-2004}
\bibfield{author}{\bibinfo{person}{Graham Cormode} {and} \bibinfo{person}{S.
  Muthukrishnan}.} \bibinfo{year}{2004}\natexlab{}.
\newblock \showarticletitle{An Improved Data Stream Summary: The Count-Min
  Sketch and Its Applications}. In \bibinfo{booktitle}{\emph{{LATIN} 2004:
  Theoretical Informatics, 6th Latin American Symposium, Buenos Aires,
  Argentina, April 5-8, 2004, Proceedings}} \emph{(\bibinfo{series}{Lecture
  Notes in Computer Science}, Vol.~\bibinfo{volume}{2976})},
  \bibfield{editor}{\bibinfo{person}{Martin Farach{-}Colton}} (Ed.).
  \bibinfo{publisher}{Springer}, \bibinfo{address}{Berlin, Germany},
  \bibinfo{pages}{29--38}.
\newblock
\urldef\tempurl%
\url{https://doi.org/10.1007/978-3-540-24698-5\_7}
\showDOI{\tempurl}


\bibitem[Cormode and Muthukrishnan(2005)]%
        {CormodeMuthukrishnan-2005}
\bibfield{author}{\bibinfo{person}{Graham Cormode} {and} \bibinfo{person}{S.
  Muthukrishnan}.} \bibinfo{year}{2005}\natexlab{}.
\newblock \showarticletitle{Space Efficient Mining of Multigraph Streams}. In
  \bibinfo{booktitle}{\emph{Proceedings of the Twenty-Fourth ACM
  SIGMOD-SIGACT-SIGART Symposium on Principles of Database Systems}}
  (Baltimore, Maryland) \emph{(\bibinfo{series}{PODS '05})}.
  \bibinfo{publisher}{Association for Computing Machinery},
  \bibinfo{address}{New York, NY, USA}, \bibinfo{pages}{271–282}.
\newblock
\showISBNx{1595930620}
\urldef\tempurl%
\url{https://doi.org/10.1145/1065167.1065201}
\showDOI{\tempurl}


\bibitem[Dasgupta et~al\mbox{.}(2016)]%
        {DLRT-2016}
\bibfield{author}{\bibinfo{person}{Anirban Dasgupta}, \bibinfo{person}{Kevin~J.
  Lang}, \bibinfo{person}{Lee Rhodes}, {and} \bibinfo{person}{Justin Thaler}.}
  \bibinfo{year}{2016}\natexlab{}.
\newblock \showarticletitle{A Framework for Estimating Stream Expression
  Cardinalities}. In \bibinfo{booktitle}{\emph{19th International Conference on
  Database Theory, {ICDT} 2016, Bordeaux, France, March 15-18, 2016}}
  \emph{(\bibinfo{series}{LIPIcs}, Vol.~\bibinfo{volume}{48})},
  \bibfield{editor}{\bibinfo{person}{Wim Martens} {and} \bibinfo{person}{Thomas
  Zeume}} (Eds.). \bibinfo{publisher}{Schloss Dagstuhl - Leibniz-Zentrum
  f{\"{u}}r Informatik}, \bibinfo{address}{Wadern, Germany},
  \bibinfo{pages}{6:1--6:17}.
\newblock
\urldef\tempurl%
\url{https://doi.org/10.4230/LIPIcs.ICDT.2016.6}
\showDOI{\tempurl}


\bibitem[Datadog(2018)]%
        {Datadog-orchestration}
\bibfield{author}{\bibinfo{person}{Datadog}.} \bibinfo{year}{2018}\natexlab{}.
\newblock \bibinfo{title}{8 emerging trends in container orchestration}.
\newblock
  \bibinfo{howpublished}{\url{https://www.datadoghq.com/container-orchestration-2018}}.
\newblock
\newblock
\shownote{Accessed: 2023-09-01}.


\bibitem[Datadog(2022)]%
        {Datadog-husky}
\bibfield{author}{\bibinfo{person}{Datadog}.} \bibinfo{year}{2022}\natexlab{}.
\newblock \bibinfo{title}{Introducing Husky, Datadog's Third-Generation Event
  Store}.
\newblock
  \bibinfo{howpublished}{\url{https://www.datadoghq.com/blog/engineering/introducing-husky/}}.
\newblock
\newblock
\shownote{Accessed: 2023-09-01}.


\bibitem[Dua and Graff(2017)]%
        {DuaGraff-2019}
\bibfield{author}{\bibinfo{person}{Dheeru Dua} {and} \bibinfo{person}{Casey
  Graff}.} \bibinfo{year}{2017}\natexlab{}.
\newblock \bibinfo{title}{{UCI} Machine Learning Repository}.
\newblock
\newblock
\urldef\tempurl%
\url{http://archive.ics.uci.edu/ml}
\showURL{%
\tempurl}


\bibitem[Estan et~al\mbox{.}(2003)]%
        {EVF-2003}
\bibfield{author}{\bibinfo{person}{Cristian Estan}, \bibinfo{person}{George
  Varghese}, {and} \bibinfo{person}{Mike Fisk}.}
  \bibinfo{year}{2003}\natexlab{}.
\newblock \showarticletitle{Bitmap Algorithms for Counting Active Flows on High
  Speed Links}. In \bibinfo{booktitle}{\emph{Proceedings of the 3rd ACM SIGCOMM
  Conference on Internet Measurement}} (Miami Beach, FL, USA)
  \emph{(\bibinfo{series}{IMC '03})}. \bibinfo{publisher}{Association for
  Computing Machinery}, \bibinfo{address}{New York, NY, USA},
  \bibinfo{pages}{153–166}.
\newblock
\showISBNx{1581137737}
\urldef\tempurl%
\url{https://doi.org/10.1145/948205.948225}
\showDOI{\tempurl}


\bibitem[Feibish et~al\mbox{.}(2017)]%
        {FABCS-2017}
\bibfield{author}{\bibinfo{person}{Shir~Landau Feibish},
  \bibinfo{person}{Yehuda Afek}, \bibinfo{person}{Anat Bremler-Barr},
  \bibinfo{person}{Edith Cohen}, {and} \bibinfo{person}{Michal Shagam}.}
  \bibinfo{year}{2017}\natexlab{}.
\newblock \showarticletitle{Mitigating DNS Random Subdomain DDoS Attacks by
  Distinct Heavy Hitters Sketches}. In \bibinfo{booktitle}{\emph{Proceedings of
  the Fifth ACM/IEEE Workshop on Hot Topics in Web Systems and Technologies}}
  (San Jose, California) \emph{(\bibinfo{series}{HotWeb '17})}.
  \bibinfo{publisher}{Association for Computing Machinery},
  \bibinfo{address}{New York, NY, USA}, Article \bibinfo{articleno}{8},
  \bibinfo{numpages}{6}~pages.
\newblock
\showISBNx{9781450355278}
\urldef\tempurl%
\url{https://doi.org/10.1145/3132465.3132474}
\showDOI{\tempurl}


\bibitem[Flajolet(1990)]%
        {Flajolet-1990}
\bibfield{author}{\bibinfo{person}{Philippe Flajolet}.}
  \bibinfo{year}{1990}\natexlab{}.
\newblock \showarticletitle{On adaptive sampling}.
\newblock \bibinfo{journal}{\emph{Computing}} \bibinfo{volume}{43},
  \bibinfo{number}{4} (\bibinfo{year}{1990}), \bibinfo{pages}{391--400}.
\newblock
\urldef\tempurl%
\url{https://doi.org/10.1007/BF02241657}
\showDOI{\tempurl}


\bibitem[Flajolet et~al\mbox{.}(2007)]%
        {FFGM-2007}
\bibfield{author}{\bibinfo{person}{Philippe Flajolet}, \bibinfo{person}{Eric
  Fusy}, \bibinfo{person}{Olivier Gandouet}, {and} \bibinfo{person}{Frédéric
  Meunier}.} \bibinfo{year}{2007}\natexlab{}.
\newblock \showarticletitle{HyperLogLog: the analysis of a near-optimal
  cardinality estimation algorithm}. In \bibinfo{booktitle}{\emph{{DMTCS}
  Proceedings vol. {AH}, 2007 Conference on Analysis of Algorithms (AofA 07)}}.
  \bibinfo{publisher}{{DMTCS}}, \bibinfo{address}{Strasbourg, France},
  \bibinfo{pages}{127--146}.
\newblock
\urldef\tempurl%
\url{https://doi.org/10.46298/dmtcs.3545}
\showDOI{\tempurl}


\bibitem[Flajolet and Martin(1983)]%
        {FlajoletMartin-1983}
\bibfield{author}{\bibinfo{person}{Philippe Flajolet} {and}
  \bibinfo{person}{G.~Nigel Martin}.} \bibinfo{year}{1983}\natexlab{}.
\newblock \showarticletitle{Probabilistic Counting}. In
  \bibinfo{booktitle}{\emph{24th Annual Symposium on Foundations of Computer
  Science, Tucson, Arizona, USA, 7-9 November 1983}}.
  \bibinfo{publisher}{{IEEE} Computer Society}, \bibinfo{address}{Piscataway,
  NJ, USA}, \bibinfo{pages}{76--82}.
\newblock
\urldef\tempurl%
\url{https://doi.org/10.1109/SFCS.1983.46}
\showDOI{\tempurl}


\bibitem[Ganguly et~al\mbox{.}(2007)]%
        {GGRS-2007}
\bibfield{author}{\bibinfo{person}{Sumit Ganguly}, \bibinfo{person}{Minos
  Garofalakis}, \bibinfo{person}{Rajeev Rastogi}, {and}
  \bibinfo{person}{Krishan Sabnani}.} \bibinfo{year}{2007}\natexlab{}.
\newblock \showarticletitle{Streaming Algorithms for Robust, Real-Time
  Detection of DDoS Attacks}. In \bibinfo{booktitle}{\emph{27th International
  Conference on Distributed Computing Systems (ICDCS '07)}}.
  \bibinfo{publisher}{{IEEE} Computer Society}, \bibinfo{address}{Piscataway,
  NJ, USA}, \bibinfo{pages}{4--4}.
\newblock
\urldef\tempurl%
\url{https://doi.org/10.1109/ICDCS.2007.142}
\showDOI{\tempurl}


\bibitem[Gormley and Tong(2015)]%
        {GormleyTong-2015}
\bibfield{author}{\bibinfo{person}{C. Gormley} {and} \bibinfo{person}{Z.
  Tong}.} \bibinfo{year}{2015}\natexlab{}.
\newblock \bibinfo{booktitle}{\emph{Elasticsearch: The Definitive Guide: A
  Distributed Real-Time Search and Analytics Engine}}.
\newblock \bibinfo{publisher}{"O'Reilly Media, Inc."},
  \bibinfo{address}{Sebastopol, CA, USA}.
\newblock


\bibitem[Heule et~al\mbox{.}(2013)]%
        {HNH-2013}
\bibfield{author}{\bibinfo{person}{Stefan Heule}, \bibinfo{person}{Marc
  Nunkesser}, {and} \bibinfo{person}{Alexander Hall}.}
  \bibinfo{year}{2013}\natexlab{}.
\newblock \showarticletitle{HyperLogLog in Practice: Algorithmic Engineering of
  a State of the Art Cardinality Estimation Algorithm}. In
  \bibinfo{booktitle}{\emph{Proceedings of the 16th International Conference on
  Extending Database Technology}} (Genoa, Italy) \emph{(\bibinfo{series}{EDBT
  '13})}. \bibinfo{publisher}{Association for Computing Machinery},
  \bibinfo{address}{New York, NY, USA}, \bibinfo{pages}{683–692}.
\newblock
\showISBNx{9781450315975}
\urldef\tempurl%
\url{https://doi.org/10.1145/2452376.2452456}
\showDOI{\tempurl}


\bibitem[Huang et~al\mbox{.}(2012)]%
        {HTY-2012}
\bibfield{author}{\bibinfo{person}{Zengfeng Huang}, \bibinfo{person}{Ke Yi},
  {and} \bibinfo{person}{Qin Zhang}.} \bibinfo{year}{2012}\natexlab{}.
\newblock \showarticletitle{Randomized algorithms for tracking distributed
  count, frequencies, and ranks}. In \bibinfo{booktitle}{\emph{Proceedings of
  the 31st {ACM} {SIGMOD-SIGACT-SIGART} Symposium on Principles of Database
  Systems, {PODS} 2012, Scottsdale, AZ, USA, May 20-24, 2012}},
  \bibfield{editor}{\bibinfo{person}{Michael Benedikt}, \bibinfo{person}{Markus
  Kr{\"{o}}tzsch}, {and} \bibinfo{person}{Maurizio Lenzerini}} (Eds.).
  \bibinfo{publisher}{{ACM}}, \bibinfo{address}{New York, NY USA},
  \bibinfo{pages}{295--306}.
\newblock
\urldef\tempurl%
\url{https://doi.org/10.1145/2213556.2213596}
\showDOI{\tempurl}
\newblock
\shownote{Full version at http://arxiv.org/abs/1412.1763}.


\bibitem[Jayaram and Woodruff(2018)]%
        {JayaramWoodruff-2018}
\bibfield{author}{\bibinfo{person}{Rajesh Jayaram} {and}
  \bibinfo{person}{David~P. Woodruff}.} \bibinfo{year}{2018}\natexlab{}.
\newblock \showarticletitle{Data Streams with Bounded Deletions}. In
  \bibinfo{booktitle}{\emph{Proceedings of the 37th ACM SIGMOD-SIGACT-SIGAI
  Symposium on Principles of Database Systems}} (Houston, TX, USA)
  \emph{(\bibinfo{series}{PODS '18})}. \bibinfo{publisher}{Association for
  Computing Machinery}, \bibinfo{address}{New York, NY, USA},
  \bibinfo{pages}{341–354}.
\newblock
\showISBNx{9781450347068}
\urldef\tempurl%
\url{https://doi.org/10.1145/3196959.3196986}
\showDOI{\tempurl}


\bibitem[Kamiyama et~al\mbox{.}(2007)]%
        {KMK-2007}
\bibfield{author}{\bibinfo{person}{N. Kamiyama}, \bibinfo{person}{T. Mori},
  {and} \bibinfo{person}{R. Kawahara}.} \bibinfo{year}{2007}\natexlab{}.
\newblock \showarticletitle{Simple and Adaptive Identification of
  Superspreaders by Flow Sampling}. In \bibinfo{booktitle}{\emph{IEEE INFOCOM
  2007 - 26th IEEE International Conference on Computer Communications}}.
  \bibinfo{publisher}{{IEEE}}, \bibinfo{address}{Piscataway, NJ, USA},
  \bibinfo{pages}{2481--2485}.
\newblock
\urldef\tempurl%
\url{https://doi.org/10.1109/INFCOM.2007.305}
\showDOI{\tempurl}


\bibitem[Kane et~al\mbox{.}(2010)]%
        {KNW-2010}
\bibfield{author}{\bibinfo{person}{Daniel~M. Kane}, \bibinfo{person}{Jelani
  Nelson}, {and} \bibinfo{person}{David~P. Woodruff}.}
  \bibinfo{year}{2010}\natexlab{}.
\newblock \showarticletitle{An optimal algorithm for the distinct elements
  problem}. In \bibinfo{booktitle}{\emph{Proceedings of the Twenty-Ninth {ACM}
  {SIGMOD-SIGACT-SIGART} Symposium on Principles of Database Systems, {PODS}
  2010, June 6-11, 2010, Indianapolis, Indiana, {USA}}},
  \bibfield{editor}{\bibinfo{person}{Jan Paredaens} {and}
  \bibinfo{person}{Dirk~Van Gucht}} (Eds.). \bibinfo{publisher}{{ACM}},
  \bibinfo{address}{New York, NY, {USA}}, \bibinfo{pages}{41--52}.
\newblock
\urldef\tempurl%
\url{https://doi.org/10.1145/1807085.1807094}
\showDOI{\tempurl}


\bibitem[Lang(2017)]%
        {Lang-2017}
\bibfield{author}{\bibinfo{person}{Kevin~J. Lang}.}
  \bibinfo{year}{2017}\natexlab{}.
\newblock \bibinfo{title}{Back to the Future: an Even More Nearly Optimal
  Cardinality Estimation Algorithm}.
\newblock
\newblock
\showeprint[arXiv]{1708.06839}
\urldef\tempurl%
\url{http://arxiv.org/abs/1708.06839}
\showURL{%
\tempurl}


\bibitem[Li et~al\mbox{.}(2013)]%
        {LCLZQ-2013}
\bibfield{author}{\bibinfo{person}{Tao Li}, \bibinfo{person}{Shigang Chen},
  \bibinfo{person}{Wen Luo}, \bibinfo{person}{Ming Zhang}, {and}
  \bibinfo{person}{Yan Qiao}.} \bibinfo{year}{2013}\natexlab{}.
\newblock \showarticletitle{Spreader Classification Based on Optimal Dynamic
  Bit Sharing}.
\newblock \bibinfo{journal}{\emph{IEEE/ACM Transactions on Networking}}
  \bibinfo{volume}{21}, \bibinfo{number}{3} (\bibinfo{year}{2013}),
  \bibinfo{pages}{817--830}.
\newblock
\urldef\tempurl%
\url{https://doi.org/10.1109/TNET.2012.2218255}
\showDOI{\tempurl}


\bibitem[Liu et~al\mbox{.}(2016b)]%
        {LQGL-2016}
\bibfield{author}{\bibinfo{person}{Weijiang Liu}, \bibinfo{person}{Wenyu Qu},
  \bibinfo{person}{Jian Gong}, {and} \bibinfo{person}{Keqiu Li}.}
  \bibinfo{year}{2016}\natexlab{b}.
\newblock \showarticletitle{Detection of Superpoints Using a Vector Bloom
  Filter}.
\newblock \bibinfo{journal}{\emph{IEEE Transactions on Information Forensics
  and Security}} \bibinfo{volume}{11}, \bibinfo{number}{3}
  (\bibinfo{year}{2016}), \bibinfo{pages}{514--527}.
\newblock
\urldef\tempurl%
\url{https://doi.org/10.1109/TIFS.2015.2503269}
\showDOI{\tempurl}


\bibitem[Liu et~al\mbox{.}(2016a)]%
        {LCG-2016}
\bibfield{author}{\bibinfo{person}{Yang Liu}, \bibinfo{person}{Wenji Chen},
  {and} \bibinfo{person}{Yong Guan}.} \bibinfo{year}{2016}\natexlab{a}.
\newblock \showarticletitle{Identifying High-Cardinality Hosts from
  Network-Wide Traffic Measurements}.
\newblock \bibinfo{journal}{\emph{IEEE Transactions on Dependable and Secure
  Computing}} \bibinfo{volume}{13}, \bibinfo{number}{5} (\bibinfo{year}{2016}),
  \bibinfo{pages}{547--558}.
\newblock
\urldef\tempurl%
\url{https://doi.org/10.1109/TDSC.2015.2423675}
\showDOI{\tempurl}


\bibitem[Locher(2011)]%
        {Locher-2011}
\bibfield{author}{\bibinfo{person}{Thomas Locher}.}
  \bibinfo{year}{2011}\natexlab{}.
\newblock \showarticletitle{Finding Heavy Distinct Hitters in Data Streams}. In
  \bibinfo{booktitle}{\emph{Proceedings of the Twenty-Third Annual ACM
  Symposium on Parallelism in Algorithms and Architectures}} (San Jose,
  California, USA) \emph{(\bibinfo{series}{SPAA '11})}.
  \bibinfo{publisher}{Association for Computing Machinery},
  \bibinfo{address}{New York, NY, USA}, \bibinfo{pages}{299–308}.
\newblock
\showISBNx{9781450307437}
\urldef\tempurl%
\url{https://doi.org/10.1145/1989493.1989541}
\showDOI{\tempurl}


\bibitem[Masson et~al\mbox{.}(2019)]%
        {MRL-2019}
\bibfield{author}{\bibinfo{person}{Charles Masson}, \bibinfo{person}{Jee~E
  Rim}, {and} \bibinfo{person}{Homin~K Lee}.} \bibinfo{year}{2019}\natexlab{}.
\newblock \showarticletitle{{DDS}ketch: A fast and fully-mergeable quantile
  sketch with relative-error guarantees}.
\newblock \bibinfo{journal}{\emph{The VLDB Journal}} \bibinfo{volume}{12},
  \bibinfo{number}{12} (\bibinfo{year}{2019}), \bibinfo{pages}{2195--2205}.
\newblock


\bibitem[Masson and Watt(2019)]%
        {MassonWatt-2019}
\bibfield{author}{\bibinfo{person}{Charles Masson} {and} \bibinfo{person}{Celia
  Watt}.} \bibinfo{year}{2019}\natexlab{}.
\newblock \bibinfo{title}{Computing Accurate Percentiles with DDSketch}.
\newblock
  \bibinfo{howpublished}{\url{https://www.datadoghq.com/blog/engineering/computing-accurate-percentiles-with-ddsketch}}.
\newblock
\newblock
\shownote{Accessed: 2023-06-27}.


\bibitem[Metwally et~al\mbox{.}(2005)]%
        {MAA-2005}
\bibfield{author}{\bibinfo{person}{Ahmed Metwally}, \bibinfo{person}{Divyakant
  Agrawal}, {and} \bibinfo{person}{Amr~El Abbadi}.}
  \bibinfo{year}{2005}\natexlab{}.
\newblock \showarticletitle{Efficient Computation of Frequent and Top-k
  Elements in Data Streams}. In \bibinfo{booktitle}{\emph{Database Theory -
  {ICDT} 2005, 10th International Conference, Edinburgh, UK, January 5-7, 2005,
  Proceedings}} \emph{(\bibinfo{series}{Lecture Notes in Computer Science},
  Vol.~\bibinfo{volume}{3363})}, \bibfield{editor}{\bibinfo{person}{Thomas
  Eiter} {and} \bibinfo{person}{Leonid Libkin}} (Eds.).
  \bibinfo{publisher}{Springer}, \bibinfo{address}{Berlin, Germany},
  \bibinfo{pages}{398--412}.
\newblock
\urldef\tempurl%
\url{https://doi.org/10.1007/978-3-540-30570-5\_27}
\showDOI{\tempurl}


\bibitem[Metwally et~al\mbox{.}(2008)]%
        {MAA-2008}
\bibfield{author}{\bibinfo{person}{Ahmed Metwally}, \bibinfo{person}{Divyakant
  Agrawal}, {and} \bibinfo{person}{Amr~El Abbadi}.}
  \bibinfo{year}{2008}\natexlab{}.
\newblock \showarticletitle{Why Go Logarithmic If We Can Go Linear? Towards
  Effective Distinct Counting of Search Traffic}. In
  \bibinfo{booktitle}{\emph{Proceedings of the 11th International Conference on
  Extending Database Technology: Advances in Database Technology}} (Nantes,
  France) \emph{(\bibinfo{series}{EDBT '08})}. \bibinfo{publisher}{Association
  for Computing Machinery}, \bibinfo{address}{New York, NY, USA},
  \bibinfo{pages}{618–629}.
\newblock
\showISBNx{9781595939265}
\urldef\tempurl%
\url{https://doi.org/10.1145/1353343.1353418}
\showDOI{\tempurl}


\bibitem[Misra and Gries(1982)]%
        {MisraGries-1982}
\bibfield{author}{\bibinfo{person}{Jayadev Misra} {and} \bibinfo{person}{David
  Gries}.} \bibinfo{year}{1982}\natexlab{}.
\newblock \showarticletitle{Finding Repeated Elements}.
\newblock \bibinfo{journal}{\emph{Sci. Comput. Program.}} \bibinfo{volume}{2},
  \bibinfo{number}{2} (\bibinfo{year}{1982}), \bibinfo{pages}{143--152}.
\newblock
\urldef\tempurl%
\url{https://doi.org/10.1016/0167-6423(82)90012-0}
\showDOI{\tempurl}


\bibitem[Plonka(2000)]%
        {Plonka-2000}
\bibfield{author}{\bibinfo{person}{Dave Plonka}.}
  \bibinfo{year}{2000}\natexlab{}.
\newblock \showarticletitle{FlowScan: A Network Traffic Flow Reporting and
  Visualization Tool}. In \bibinfo{booktitle}{\emph{Proceedings of the 14th
  USENIX Conference on System Administration}} (New Orleans, Louisiana)
  \emph{(\bibinfo{series}{LISA '00})}. \bibinfo{publisher}{USENIX Association},
  \bibinfo{address}{USA}, \bibinfo{pages}{305–318}.
\newblock


\bibitem[Roesch(1999)]%
        {Roesch-1999}
\bibfield{author}{\bibinfo{person}{Martin Roesch}.}
  \bibinfo{year}{1999}\natexlab{}.
\newblock \showarticletitle{Snort - Lightweight Intrusion Detection for
  Networks}. In \bibinfo{booktitle}{\emph{Proceedings of the 13th USENIX
  Conference on System Administration}} (Seattle, Washington)
  \emph{(\bibinfo{series}{LISA '99})}. \bibinfo{publisher}{USENIX Association},
  \bibinfo{address}{USA}, \bibinfo{pages}{229–238}.
\newblock


\bibitem[Shahout et~al\mbox{.}(2022)]%
        {SFB-2022}
\bibfield{author}{\bibinfo{person}{Rana Shahout}, \bibinfo{person}{Roy
  Friedman}, {and} \bibinfo{person}{Ran~Ben Basat}.}
  \bibinfo{year}{2022}\natexlab{}.
\newblock \showarticletitle{SQUAD: Combining Sketching and Sampling is Better
  than Either for per-Item Quantile Estimation}. In
  \bibinfo{booktitle}{\emph{Proceedings of the 15th ACM International
  Conference on Systems and Storage}} (Haifa, Israel)
  \emph{(\bibinfo{series}{SYSTOR '22})}. \bibinfo{publisher}{Association for
  Computing Machinery}, \bibinfo{address}{New York, NY, USA},
  \bibinfo{pages}{152}.
\newblock
\showISBNx{9781450393805}
\urldef\tempurl%
\url{https://doi.org/10.1145/3534056.3535009}
\showDOI{\tempurl}


\bibitem[Sidana et~al\mbox{.}(2017)]%
        {SLAVB-2017}
\bibfield{author}{\bibinfo{person}{Sumit Sidana}, \bibinfo{person}{Charlotte
  Laclau}, \bibinfo{person}{Massih~R. Amini}, \bibinfo{person}{Gilles
  Vandelle}, {and} \bibinfo{person}{Andr\'{e} Bois-Crettez}.}
  \bibinfo{year}{2017}\natexlab{}.
\newblock \showarticletitle{KASANDR: A Large-Scale Dataset with Implicit
  Feedback for Recommendation}. In \bibinfo{booktitle}{\emph{Proceedings of the
  40th International ACM SIGIR Conference on Research and Development in
  Information Retrieval}} (Shinjuku, Tokyo, Japan)
  \emph{(\bibinfo{series}{SIGIR '17})}. \bibinfo{publisher}{Association for
  Computing Machinery}, \bibinfo{address}{New York, NY, USA},
  \bibinfo{pages}{1245–1248}.
\newblock
\showISBNx{9781450350228}
\urldef\tempurl%
\url{https://doi.org/10.1145/3077136.3080713}
\showDOI{\tempurl}


\bibitem[Tang et~al\mbox{.}(2020)]%
        {THL-2020}
\bibfield{author}{\bibinfo{person}{Lu Tang}, \bibinfo{person}{Qun Huang}, {and}
  \bibinfo{person}{Patrick P.~C. Lee}.} \bibinfo{year}{2020}\natexlab{}.
\newblock \showarticletitle{SpreadSketch: Toward Invertible and Network-Wide
  Detection of Superspreaders}. In \bibinfo{booktitle}{\emph{IEEE INFOCOM 2020
  - IEEE Conference on Computer Communications}}. \bibinfo{publisher}{{IEEE}},
  \bibinfo{address}{Piscataway, NJ, USA}, \bibinfo{pages}{1608--1617}.
\newblock
\urldef\tempurl%
\url{https://doi.org/10.1109/INFOCOM41043.2020.9155541}
\showDOI{\tempurl}


\bibitem[Ting(2019)]%
        {Ting-2019}
\bibfield{author}{\bibinfo{person}{Daniel Ting}.}
  \bibinfo{year}{2019}\natexlab{}.
\newblock \showarticletitle{Approximate Distinct Counts for Billions of
  Datasets}. In \bibinfo{booktitle}{\emph{Proceedings of the 2019 International
  Conference on Management of Data, {SIGMOD} Conference 2019, Amsterdam, The
  Netherlands, June 30 - July 5, 2019}},
  \bibfield{editor}{\bibinfo{person}{Peter~A. Boncz}, \bibinfo{person}{Stefan
  Manegold}, \bibinfo{person}{Anastasia Ailamaki}, \bibinfo{person}{Amol
  Deshpande}, {and} \bibinfo{person}{Tim Kraska}} (Eds.).
  \bibinfo{publisher}{{ACM}}, \bibinfo{address}{New York, {NY}, {USA}},
  \bibinfo{pages}{69--86}.
\newblock
\urldef\tempurl%
\url{https://doi.org/10.1145/3299869.3319897}
\showDOI{\tempurl}


\bibitem[Venkataraman et~al\mbox{.}(2005)]%
        {VSGB-2005}
\bibfield{author}{\bibinfo{person}{Shobha Venkataraman}, \bibinfo{person}{Dawn
  Song}, \bibinfo{person}{Phillip~B Gibbons}, {and} \bibinfo{person}{Avrim
  Blum}.} \bibinfo{year}{2005}\natexlab{}.
\newblock \showarticletitle{New streaming algorithms for fast detection of
  superspreaders}. In \bibinfo{booktitle}{\emph{Proceedings of the Network and
  Distributed System Security Symposium, NDSS 2005, San Diego, California,
  {USA}}}. \bibinfo{publisher}{The Internet Society}, \bibinfo{address}{Reston,
  VA}, \bibinfo{pages}{1–--1}.
\newblock


\bibitem[Wang et~al\mbox{.}(2011)]%
        {WGQH-2011}
\bibfield{author}{\bibinfo{person}{Pinghui Wang}, \bibinfo{person}{Xiaohong
  Guan}, \bibinfo{person}{Tao Qin}, {and} \bibinfo{person}{Qiuzhen Huang}.}
  \bibinfo{year}{2011}\natexlab{}.
\newblock \showarticletitle{A Data Streaming Method for Monitoring Host
  Connection Degrees of High-Speed Links}.
\newblock \bibinfo{journal}{\emph{IEEE Transactions on Information Forensics
  and Security}} \bibinfo{volume}{6}, \bibinfo{number}{3}
  (\bibinfo{year}{2011}), \bibinfo{pages}{1086--1098}.
\newblock
\urldef\tempurl%
\url{https://doi.org/10.1109/TIFS.2011.2123094}
\showDOI{\tempurl}


\bibitem[Weber(2014)]%
        {Weber-2014}
\bibfield{author}{\bibinfo{person}{Ralf Weber}.}
  \bibinfo{year}{2014}\natexlab{}.
\newblock \bibinfo{title}{Latest Internet Plague: Random Subdomain Attacks}.
  (\bibinfo{year}{2014}).
\newblock
\urldef\tempurl%
\url{https://indico.uknof.org.uk/event/31/contributions/349/}
\showURL{%
\tempurl}
\newblock
\shownote{UKNOF29 \& Internet Society ION Conference}.


\bibitem[Xiao et~al\mbox{.}(2015)]%
        {XCCL-2015}
\bibfield{author}{\bibinfo{person}{Qingjun Xiao}, \bibinfo{person}{Shigang
  Chen}, \bibinfo{person}{Min Chen}, {and} \bibinfo{person}{Yibei Ling}.}
  \bibinfo{year}{2015}\natexlab{}.
\newblock \showarticletitle{Hyper-Compact Virtual Estimators for Big Network
  Data Based on Register Sharing}. In \bibinfo{booktitle}{\emph{Proceedings of
  the 2015 {ACM} {SIGMETRICS} International Conference on Measurement and
  Modeling of Computer Systems, Portland, OR, USA, June 15-19, 2015}},
  \bibfield{editor}{\bibinfo{person}{Bill Lin}, \bibinfo{person}{Jun~(Jim) Xu},
  \bibinfo{person}{Sudipta Sengupta}, {and} \bibinfo{person}{Devavrat Shah}}
  (Eds.). \bibinfo{publisher}{{ACM}}, \bibinfo{address}{New York, NY, USA},
  \bibinfo{pages}{417--428}.
\newblock
\urldef\tempurl%
\url{https://doi.org/10.1145/2745844.2745870}
\showDOI{\tempurl}


\bibitem[Yoon et~al\mbox{.}(2011)]%
        {YLCP-2011}
\bibfield{author}{\bibinfo{person}{MyungKeun Yoon}, \bibinfo{person}{Tao Li},
  \bibinfo{person}{Shigang Chen}, {and} \bibinfo{person}{Jih-Kwon Peir}.}
  \bibinfo{year}{2011}\natexlab{}.
\newblock \showarticletitle{Fit a Compact Spread Estimator in Small High-Speed
  Memory}.
\newblock \bibinfo{journal}{\emph{IEEE/ACM Trans. Netw.}} \bibinfo{volume}{19},
  \bibinfo{number}{5} (\bibinfo{date}{oct} \bibinfo{year}{2011}),
  \bibinfo{pages}{1253–--1264}.
\newblock
\showISSN{1063-6692}
\urldef\tempurl%
\url{https://doi.org/10.1109/TNET.2010.2080285}
\showDOI{\tempurl}


\bibitem[Yu et~al\mbox{.}(2013)]%
        {YJM-2013}
\bibfield{author}{\bibinfo{person}{Minlan Yu}, \bibinfo{person}{Lavanya Jose},
  {and} \bibinfo{person}{Rui Miao}.} \bibinfo{year}{2013}\natexlab{}.
\newblock \showarticletitle{Software Defined Traffic Measurement with
  OpenSketch}. In \bibinfo{booktitle}{\emph{Proceedings of the 10th USENIX
  Conference on Networked Systems Design and Implementation}} (Lombard, IL)
  \emph{(\bibinfo{series}{NSDI '13})}. \bibinfo{publisher}{USENIX Association},
  \bibinfo{address}{USA}, \bibinfo{pages}{29–42}.
\newblock


\bibitem[Zhao et~al\mbox{.}(2022)]%
        {ZAAM-2022}
\bibfield{author}{\bibinfo{person}{Fuheng Zhao}, \bibinfo{person}{Divyakant
  Agrawal}, \bibinfo{person}{Amr~El Abbadi}, {and} \bibinfo{person}{Ahmed
  Metwally}.} \bibinfo{year}{2022}\natexlab{}.
\newblock \showarticletitle{SpaceSaving±: An Optimal Algorithm for Frequency
  Estimation and Frequent Items in the Bounded-Deletion Model}.
\newblock \bibinfo{journal}{\emph{Proc. VLDB Endow.}} \bibinfo{volume}{15},
  \bibinfo{number}{6} (\bibinfo{date}{feb} \bibinfo{year}{2022}),
  \bibinfo{pages}{1215–1227}.
\newblock
\showISSN{2150-8097}
\urldef\tempurl%
\url{https://doi.org/10.14778/3514061.3514068}
\showDOI{\tempurl}


\bibitem[Zhao et~al\mbox{.}(2021)]%
        {ZMWAA-2021}
\bibfield{author}{\bibinfo{person}{Fuheng Zhao}, \bibinfo{person}{Sujaya
  Maiyya}, \bibinfo{person}{Ryan Wiener}, \bibinfo{person}{Divyakant Agrawal},
  {and} \bibinfo{person}{Amr~El Abbadi}.} \bibinfo{year}{2021}\natexlab{}.
\newblock \showarticletitle{KLL± Approximate Quantile Sketches over Dynamic
  Datasets}.
\newblock \bibinfo{journal}{\emph{Proc. VLDB Endow.}} \bibinfo{volume}{14},
  \bibinfo{number}{7} (\bibinfo{date}{mar} \bibinfo{year}{2021}),
  \bibinfo{pages}{1215–1227}.
\newblock
\showISSN{2150-8097}
\urldef\tempurl%
\url{https://doi.org/10.14778/3450980.3450990}
\showDOI{\tempurl}


\bibitem[Zhao et~al\mbox{.}(2005)]%
        {ZKX-2005}
\bibfield{author}{\bibinfo{person}{Qi Zhao}, \bibinfo{person}{Abhishek Kumar},
  {and} \bibinfo{person}{Jun Xu}.} \bibinfo{year}{2005}\natexlab{}.
\newblock \showarticletitle{Joint Data Streaming and Sampling Techniques for
  Detection of Super Sources and Destinations}. In
  \bibinfo{booktitle}{\emph{Proceedings of the 5th ACM SIGCOMM Conference on
  Internet Measurement}} (Berkeley, CA) \emph{(\bibinfo{series}{IMC '05})}.
  \bibinfo{publisher}{USENIX Association}, \bibinfo{address}{USA},
  \bibinfo{pages}{7}.
\newblock


\end{thebibliography}
\balance

\end{document}